\documentclass{aa}

\usepackage{graphicx}
\usepackage[colorlinks=true, linkcolor=blue, citecolor=blue]{hyperref}
\newcommand{\tiu}{J~m$^{-2}$~K$^{-1}$~s$^{-1/2}$}

\usepackage{txfonts}

\providecommand\linenumbers{}     
\providecommand\nolinenumbers{}   
\makeatletter
\@ifundefined{nolinenumbers}{}{
    \nolinenumbers
}
\makeatother

\begin{document} 

\makeatletter
\def\linenumbers{\relax}
\def\thelinenumber{}
\makeatother
\title{Thermophysical properties of Europa's surface constrained by Galileo photopolarimeter-radiometer temperature measurements}

   \author{L. Lange \inst{1}, S. Piqueux \inst{2}, P. O. Hayne \inst{3}, C. Mergny\inst{4,5}, A. Le Gall\inst{6}, F. Schmidt\inst{4,7},   J. Rathbun\inst{8}, J. Spencer\inst{9}, T. Nordheim\inst{10}, K. Sorli\inst{11}, S. Howes\inst{11}, C. Howett\inst{11}, C.S. Edwards\inst{12}, P.R. Christensen\inst{13}}
   \titlerunning{Thermophysical properties of Europa's surface constrained by Galileo PPR temperature measurements}

\authorrunning{Lange et al.}

   \institute{NASA Postdoctoral Program Fellow at Jet Propulsion Laboratory (JPL), California Institute of Technology, Pasadena, CA, USA.\\
              \email{lucas.lange@jpl.nasa.gov}
         \and
           Jet Propulsion Laboratory (JPL), California Institute of Technology, Pasadena, CA, USA.
           \and  University of Colorado, Boulder, CO, USA.
\and  University Paris-Saclay, GEOPS, UMR 8148, CNRS, Orsay, France.
\and European Space Agency (ESA), European Space Astronomy Center (ESAC), Villanueva de la Cañada, Spain.
\and LATMOS/IPSL, Sorbonne Université, UVSQ, CNRS, Paris, France
\and Institut Universitaire de France, IUF
\and  Cornell University, Ithaca, NY, USA.
\and  Southwest Research Institute, Boulder, CO, USA.
\and Johns Hopkins Applied Physics Laboratory, Laurel, MD, USA.
\and  University of Oxford, Oxford, UK.
\and  Northern Arizona University, Flagstaff, AZ, USA.
\and  Arizona State University, Tempe, AZ, USA}

  \abstract
{Thermal measurements provide key constraints on the physical properties of icy satellite surfaces. On Europa, previous analyses of the Galileo Photopolarimeter-Radiometer (PPR) dataset revealed heterogeneities in thermal inertia, but limited spatial resolution prevented a detailed thermophysical characterization.}
{We aim to derive high-resolution maps of Europa's surface albedo and thermal inertia, and to infer the microphysical properties of its icy regolith from a reanalysis of the Galileo PPR dataset, while discussing the processes controlling its thermophysical evolution.}
{We use the KRC thermal model (K referring to the  conductivity $\kappa$, R to the density $\rho$, and C to the specific heat $C$) to analyze the PPR brightness temperatures and retrieve the albedo and thermal inertia. These values are then interpreted using conductivity models of porous ice to constrain grain size and porosity.}
{We derive a mean Bond albedo of $0.64~\pm~0.06$ and a mean thermal inertia of $56~\pm~17$~J~m$^{-2}$~K$^{-1}$~s$^{-1/2}$~($1\sigma$). The thermal inertia shows significant spatial variations, with a low-inertia equatorial band ($39~\pm~7$) and higher values at mid-latitudes on the leading hemisphere ($56~\pm~11$). The trailing-hemisphere equator also exhibits higher thermal inertia ($63~\pm~17$), likely related to compositional differences. Conductivity models indicate a porous icy regolith with grain sizes ranging from a few micrometers to a few centimeters and an average porosity of $0.61~\pm~0.1$ in the upper centimeters of Europa's surface.}
{The thermal inertia distribution shows little correlation with geological units. Its agreement with modeled magnetospheric ion fluxes instead suggests that sputtering-driven sintering plays a fundamental role in shaping Europa's thermophysical properties. The absence of a high-inertia equatorial band analogous to the PacMan anomaly on Saturn's icy moons indicates inefficient electron-driven sintering, while temperature-gradient metamorphism may enhance grain growth at depth. Modeled surface temperatures range between $\sim$~67 and 148 K at mid to low latitudes, with peak daytime values counteracting radiolytic amorphization while limiting volatile stability.}

   \keywords{Planets and satellites: individual: Europa; Planets and satellites: surfaces; Methods: numerical;  Methods: observational.  }

\maketitle

\section{Introduction}
Europa, the smallest of Jupiter's four Galilean moons, is widely regarded as one of the most promising environments in the Solar System for the emergence of extraterrestrial life. Multiple lines of evidence point to the presence of a global salty ocean beneath its icy surface \citep{Khurana1998,Carr1998,Pappalardo1998}, potentially containing the chemical ingredients and energy sources required to sustain life \citep[see review in][]{Vance2023}. Surface features such as chaos terrains, lenticulae, and double ridges observed on Europa's relatively young surface suggest the presence of liquid brine reservoirs within the shallow ice shell \citep{Pappalardo1998,Schmidt2011,Culberg2022}. These subsurface liquid reservoirs may occasionally breach the surface \citep{Lesage_SimulationFreezingCryomagma_TPSJ2022} that may form geyser-like plumes, providing a unique opportunity to directly sample material originating from the moon's interior. Such a possibility has motivated numerous observational searches for plume activity. Although several studies have reported evidence for large gaseous plumes at various locations on Europa \citep[e.g.,][]{Roth2014,Sparks2016}, no definitive conclusion has yet been reached regarding the existence or persistence of active plumes \citep{Roth2025,Roth2026}. Consequently, the detection and characterization of such plumes remain among the primary scientific objectives of the upcoming NASA Europa Clipper \citep{Pappalardo2024} and ESA JUICE \citep{Grasset2013} missions.

The presence of near-surface liquid reservoirs could be detected through the appearance of anomalously warm regions (also referred to as “hot spots”), as they would transport warm material from depth toward the near-surface \citep{Abramov2008,Abramov2013}. While such hot spots have been observed on other icy moons, such as Io \citep{Spencer2000} and Enceladus \citep{Spencer2006}, no clear thermal anomalies have been identified on Europa, even at locations where plume activity has been proposed \citep{Spencer1999,Rathbun2010,Trumbo2017,Rathbun2020}. Indeed, surface temperature measurements, whether obtained from orbit around Jupiter or from ground-based telescopes, are consistent with passive heating and cooling driven by the diurnal and seasonal solar illumination cycle \citep[see review in][]{Thelen2024}. Therefore, if thermal anomalies are present on Europa, they are likely to be subtle and/or small in size and therefore challenging to detect.

Lessons learned from past missions throughout the solar system indicate that anomalous warm surfaces at night can result from surface properties rather than from an endogenic heat source. For instance, on Mars, warm nighttime temperatures can be explained by high-conductivity cemented material, limiting nighttime cooling after sunset \citep{Christensen2003}. Similarly, high nighttime temperature observed on the bright ejecta blanket of the crater Pwyll on Europa can be explained by regional variations of the surface properties \cite{Spencer1999}. Determining these properties is hence crucial to predict surface temperatures and therefore detect accurately hotspot on Europa's surface.

Three physical properties play a key role in controlling Europa's surface temperature: the albedo,  by determining the fraction of incident solar energy absorbed and thus regulating daytime temperatures; the emissivity, controlling radiative cooling to space; and the thermal inertia, which depends on the thermal conductivity, density, and porosity of the surface material and primarily governs nighttime temperatures. The latter is particularly important because nighttime temperatures are generally preferred for thermophysical studies and for the detection of potential hot spots, as they minimize the effects of surface geometry and roughness on the thermal signal. Moreover, since thermal inertia is directly linked to the microphysical properties of the ice \citep[e.g., grain size and porosity][]{Ferrari2016}, it can be used to infer these characteristics and constrain the surface processes at work. It therefore constrains the microphysical state of the surface ice (grain size and porosity), informing models of sintering-driven grain growth \citep[e.g., ][]{Mergny2024lunaicy} and space weathering processes, including charged-particle bombardment \citep[e.g., ][]{Cooper2001} and impact gardening \citep[e.g., ][]{Zahnle1998}.

To date, the most comprehensive thermal inertia coverage of Europa's surface has been obtained from temperature measurements acquired with the Atacama Large Millimeter Array (ALMA) \citep{Trumbo2017,Trumbo2018,Thelen2024}. These studies suggest that the porosity of the upper $\sim$20 cm of Europa's subsurface ranges between 40\% and 70\%, and that the thermal inertia varies between 50 and 180~\tiu. However, these analyses relied only on daytime temperature measurements at relatively coarse spatial resolution \citep[approximately 500~km,][]{Thelen2024}, limiting the ability to correlate the derived thermophysical properties with specific geological features \citep{Leonard2024}. Moreover, because these measurements were obtained during the daytime, when surface temperatures are strongly influenced by albedo variations and surface roughness, they are not optimal for deriving thermal inertia. Nighttime observations at higher spatial resolution would thus better constrain these properties.

    The Galileo PPR dataset \citep{Russell1992} offers such measurements, with intrinsic spatial resolutions reaching $\sim$100~km/pixel and coverage of both daytime and nighttime conditions. \cite{Rathbun2010} and \cite{Rathbun2020} exploited 15 high-spatial-resolution observation sequences from this dataset, covering nearly 20\% of Europa's surface. However, these observations were subsequently binned into 10\textdegree~$\times$~10\textdegree~latitude–longitude cells (corresponding to $\sim$~300~km at the equator) to improve signal-to-noise. In addition, insolation was not corrected for latitude, potentially leading to an overestimation of solar heating at high latitudes and introducing biases in the derived surface energy budget and thermophysical properties.  A recent reanalysis of the Galileo PPR dataset was conducted by \cite{Howes2025}. Their study focuses on the detection of endogenic hotspots and on the derivation of albedo and thermal inertia maps at 6\textdegree~$\times$~6\textdegree~resolution.

In this work, we also revisit the Galileo PPR dataset at the highest resolution studied so far (0.5\textdegree~$\times$~0.5\textdegree) and with an improved thermophysical model. Our specific objectives are threefold:
\begin{enumerate}
    \item Provide revised albedo and thermal inertia maps of Europa's surface at an unprecedented resolution;
    \item Derive constraints on the porosity and grain size of the regolith from these maps;
    \item Investigate correlations between thermophysical properties and geological features/surface processes using our highest-resolution coverage.

\end{enumerate}

The data, methods, and thermophysical model used are described in Section~\ref{sec:Methods}. The resulting albedo and thermal inertia maps, along with the inferred microphysical properties of the ice (grain size and porosity), are presented in Section~\ref{sec:Results}. These findings are discussed in Section~\ref{sec:Discussion}, and conclusions are provided in Section~\ref{sec:Cl}.

\section{Data, model, and methods \label{sec:Methods}}

\subsection{Galileo PPR temperatures \label{ssec:PPR}}

We analyzed surface temperature measurements acquired by the  Photopolarimeter–Radiometer (PPR) instrument onboard Galileo between November 1996 and November 1999. Detailed descriptions of the instrument and dataset are provided in 
\cite{Russell1992}, \cite{Spencer1999}, \cite{Rathbun2004}, and \cite{Rathbun2010}. We used the same subset of observations as \cite{Rathbun2010} and \cite{Rathbun2014}, namely the sequences listed in Table~1 of \cite{Rathbun2010}, supplemented with E4DGTM01, E4GLOBAL01, E4DRKMAP02, G7DGTM\_\_02, E14DRKMAP01, E14DARKHR01, E15DRKMAP01, E15DRTMHR01, E17DRTM\_\_01, available from the Planetary Data System (PDS) \citep{GalileoPDS}.  This subset comprised 23 observation sequences out of the 137 available on the PDS and was selected because it provided the broadest spatial coverage and the highest signal-to-noise ratio. Radiometric measurements were acquired either without a spectral filter (i.e., spanning both visible and infrared wavelengths) or within channels centered at 16.8, 27.5, and 35.5~$\mu$m, each with a spectral bandwidth of approximately 4 to 7~$\mu$m.

\begin{figure}[h!] \centering \includegraphics[width=0.5\textwidth]{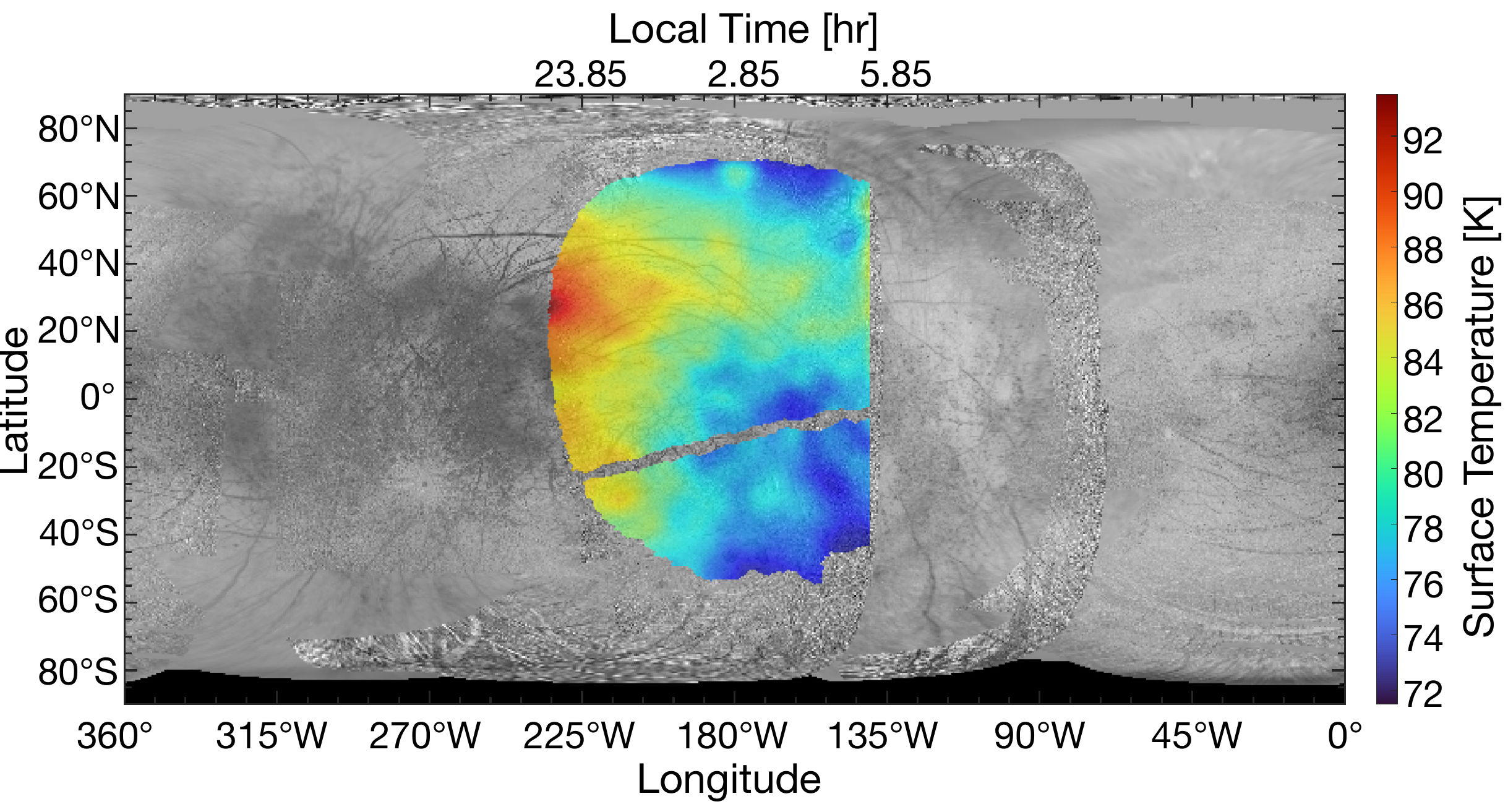} \caption{Illustration of a temperature map generated using PPR measurements from the sequence I25DRKMAP01. Background is the USGS Europa Voyager-Galileo SSI Global Mosaic of Europa \citep{Becker2010}.
      } 
    \label{fig:tempmapEuropa}
    \end{figure}

Measurements potentially contaminated by the spacecraft boom crossing the PPR field of view were discarded. Measurements acquired during Jupiter eclipses  were also removed. To limit geometric effects associated with surface roughness and illumination conditions, we excluded observations acquired near sunrise and sunset (between 5~a.m. and 10~a.m., and between 5~p.m. and 7~p.m.), as well as measurements obtained at emission angles greater than 60\textdegree. Surface temperature maps were generated at a spatial resolution of 0.5\textdegree~$\times$~0.5\textdegree~($\sim$13.6~km at the equator), following the data reduction procedure described in Section~2.1 of \cite{Rathbun2004}. In particular, despite the relatively large instantaneous field of view of the instrument (2.5~mrad), successive PPR scans overlap spatially. A spherical projection of these overlapping measured radiances, is therefore performed to reconstruct temperature maps at higher effective spatial resolution than that of the individual measurements. Additional details on the calibration procedure used to convert radiance to brightness temperature are provided in \cite{Russell1992,Spencer1999,Rathbun2004,GalileoPDS}. An example of a generated temperature map is presented in Figure~\ref{fig:tempmapEuropa}. Finally, we retained only locations with both daytime and nighttime measurements, as both are required to constrain surface thermophysical properties (Section~\ref{ssec:constrainAlbTI}). Because of these criteria, although we used a larger dataset than previous studies \citep{Spencer1999,Rathbun2010}, our spatial coverage remains similar to that of \cite{Rathbun2010}. However, our analysis achieves a substantially higher spatial resolution  (0.5\textdegree~here compared to 10\textdegree~in their study).  In addition, we cover a significant fraction of the trailing hemisphere that was not examined in the dataset of \cite{Rathbun2010}. The final dataset contains 1,161,585 temperatures, ranging from 69.9~K to 140.1~K. A conservative 2~K uncertainty on the surface temperature was assumed for every measurements \citep{Russell1992,Rathbun2020}. 

\cite{Howes2025} used a similar approach to ours. However, they included more sequences and applied looser constraints on the local time of acquisition. As a result, they cover slightly more of Europa's surface (33\% vs. 22\% in this study), while covering roughly the same areas. However, their dataset has a resolution of 6\textdegree~$\times$~6\textdegree, while ours has a resolution of 0.5\textdegree~$\times$~0.5\textdegree. Hence, our study represents the highest-resolution mapping of Europa's thermophysical properties (albedo and thermal inertia retrieved simultaneously) achieved so far.

\subsection{Thermal modeling with KRC}
To model the surface temperatures of Europa and derive the thermophysical properties of its surface, we used the  KRC thermal model  \citep{Kieffer2013}. KRC computes surface temperatures based on the balance between absorbed solar flux, thermal emission to space, and conduction into the subsurface:

\begin{equation}
    \epsilon \sigma_{\rm{SB}} T_{\rm{surf}}^4~=~ (1-A) F_{\rm{solar}} + I\sqrt{\frac{\pi}{P}}\frac{\partial T}{\partial z}
    \label{eq:GeneralTsurf}
\end{equation}

\noindent where $\epsilon$ (unitless) is the surface emissivity, $\sigma_{\rm SB} = 5.67 \times 10^{-8}$ W~m$^{-2}$~K$^{-4}$ is the Stefan–Boltzmann constant, $T_{\rm surf}$ (K) is the surface temperature, $A$ (unitless) is the Bond albedo (hereafter referred to as "albedo"), $F_{\rm solar}$ (W~m$^{-2}$) is the incident solar flux at the surface (accounting for solar incidence angle), $I$ (\tiu) is the thermal inertia, $P$ (s) is the solar forcing period~(approximatively 3.55 Earth days, i.e., 306720~s), and $T$ (K) is the subsurface temperature at depth $z$~(unitless), with $z$ being the dimensionless depth normalized by the diurnal thermal skin depth.

Here, we assume that all solar energy is absorbed at the surface and not distributed within the subsurface, i.e., no ice transparency is considered. This assumption is further discussed in Section \ref{ssec:solarpen}. We also neglect the contribution from visible light reflected by Jupiter and its infrared emission toward Europa's surface. As noted by \cite{Spencer1999}, these contributions affect surface temperatures by less than 1~K, below the uncertainty of the PPR measurements assumed here.

In our approach, albedo $A$ and thermal inertia $I$ are treated as free parameters, while the emissivity $\epsilon$ is fixed at 0.9, following \citet{Spencer1999}, who showed that Voyager thermal emission spectra of Europa are consistent with gray-body emission with a mean emissivity near 0.9. Sensitivity tests performed for emissivities ranging from 0.8 to 1.0 indicate that the resulting variations in retrieved albedo and thermal inertia remain typically within $\pm$~10\%, generally below the uncertainties associated with the PPR measurements. Finally, $I$ is assumed to be constant with depth.

\subsection{Deriving the albedo and thermal inertia of the surface \label{ssec:constrainAlbTI}}
\begin{figure}[h!] \centering \includegraphics[width=0.5\textwidth]{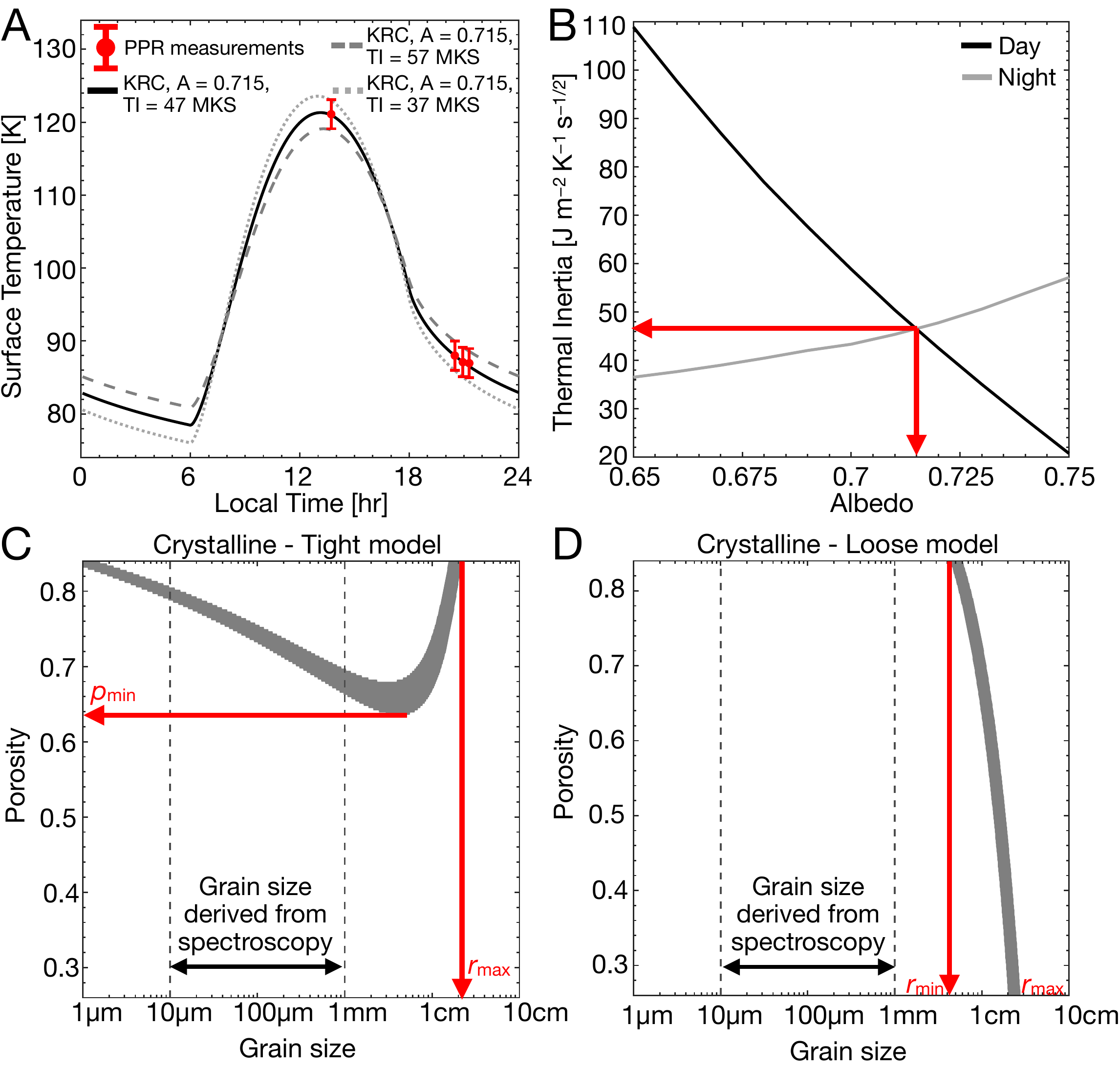} \caption{Illustration of the retrieval procedure for albedo, thermal inertia,  and the microphysical properties of the ice. a) PPR-measured surface temperatures (red points) at latitude 0.1\textdegree N\ and longitude 30\textdegree~W, with uncertainties of  $\pm$2~K. b) Model-generated albedo and thermal inertia solutions derived from a pair of daytime ($\sim$13.7~hr) and nighttime ($\sim$21.3~hr) temperatures. A unique solution albedo = 0.715 and thermal inertia = 47~\tiu (unit we abbreviate as MKS in the figure) simultaneously matches both the selected daytime (black) and  nighttime (grey) temperatures. This pair of values produces the black curve shown in panel a). Temperature curves computed using the thermal inertia values allowed by the $\pm$~2~K uncertainties are shown as grey dashed and dotted lines in panel a) Similar curves are also generated using the albedo values allowed by the $\pm$~2~K uncertainties, but not shown for the clarity of the figure. c,d) Grain size and porosity calculations consistent with the derived thermal inertia (and its uncertainty), shown in grey, for the tight (c) and loose (d) packing models assuming crystalline ice. Results for  amorphous ice are presented in Figure~\ref{fig:illustrationmethod_amorphous}. The minimum and maximum  grain sizes ($r_{\rm min}$ and $r_{\rm max}$), as well as the minimum  porosity ($p_{\rm min}$), are then extracted.  
      } 
    \label{fig:illustrationmethod}
    \end{figure}

The albedo and thermal inertia are derived from PPR measurements using the KRC thermal model, following the methodology proposed by \cite{Piqueux2021}. For each point of the surface temperature maps described in Section~\ref{ssec:PPR} (e.g., Figure~\ref{fig:illustrationmethod}a), we construct all possible pairs of observations in which the first element is acquired during the day and the second at night. For each of such pair, the albedo and thermal inertia are retrieved as follows. For both daytime and nighttime observations separately, we invert a model-generated lookup table produced by KRC over a prescribed range of albedo (0.1–0.95) and soil thermal inertia (5–500~\tiu), in order to identify the families of albedo/thermal inertia combinations capable of reproducing the observed temperatures (e.g., Figure~\ref{fig:illustrationmethod}b). Note that KRC is run for several jovian years until thermal equilibrium (at the surface and in the subsurface) is reached \citep{Kieffer2013}. The unique albedo/thermal inertia solution that simultaneously satisfies both the daytime and nighttime observations is then identified. The quality of the fit is assessed by computing the root mean square (RMS) deviation between the observed temperatures and the best-fitting model-generated diurnal/seasonal curve. In addition to this best-fit solution, we also determine the minimum (largest diurnal amplitude) and maximum (smallest diurnal amplitude) allowable albedo and thermal inertia values consistent with the surface temperature uncertainties described in Section~\ref{ssec:PPR} (e.g., Figure~\ref{fig:illustrationmethod}a). Once the albedo and thermal inertia have been derived for each daytime/nighttime observation pair, the final surface albedo and thermal inertia are computed as the median values over the full set of retrieved solutions. The uncertainty on these quantities is estimated at the 3$\sigma$ level as the maximum difference between the median values and the albedo/thermal inertia solutions derived at the boundaries of the measurement error bars.

By construction of our dataset, the observation pairs used in this analysis are generally acquired near local noon and local midnight, corresponding precisely to the times most diagnostic of thermophysical properties and  therefore well suited to this retrieval method. As noted by \cite{Piqueux2021}, using observation pairs that span a large fraction of the diurnal temperature amplitude, rather than fitting the complete diurnal cycle  provides strong constraints on both the apparent diurnal thermal inertia and the thermometric albedo. Notably, crossover times that are least diagnostic of thermophysical properties (near 7.0 and 15.0~Local True Solar Time (LTST)) are either excluded or their influence is mitigated. Furthermore, periods of high solar incidence (near 6.0~LTST and in the hours following 18.0~LTST) are not considered, as temperatures become strongly influenced by surface roughness and heterogeneity as the Sun approaches the horizon. Finally, lookup tables encompassing both diurnal and seasonal temperature variations allow correction for seasonal effects, an important consideration given that the PPR observations used here span approximately two years, during which the heliocentric distance varies by about 0.18~astronomical unit.

\subsection{Deriving the grain size and porosity of the surface \label{ssec:derivingrandp}}
The thermal inerta $I$ is defined by :
\begin{equation}
    I = \sqrt{k_{\rm{eff, ice}}(p,R,T)\times (1-p) \rho_{\rm{ice}}(T)\times c_p(T) }
    \label{eq:I}
\end{equation}
\noindent where $k_{\rm eff,ice}(p,R,T)$ (W m$^{-1}$ K$^{-1}$) is the effective thermal conductivity of the ice, $p$ (unitless) is the porosity, $R$ (m) is the grain size, $T$ (K) is the temperature, $\rho$ (kg m$^{-3}$) is the density, and $c_p$ (J kg$^{-1}$ K$^{-1}$) is the specific heat capacity. The bulk conductivity of pure water ice depends on its crystallinity and on the quality of contact between grains, which controls the contact (solid-state) conductivity, as well as on radiative conductivity within the porous medium \citep{Ferrari2016}.

We will first assume that the (sub)surface probed in this study is composed of pure crystalline water ice. Thermal inertia values previously derived for Europa \citep[see Table 1 of][]{Thelen2024}, as well as those obtained in this study, typically range between 10 and 200~\tiu, corresponding to a thermal skin depth of $\sim$3~mm to $\sim$6~cm. Spectroscopic studies \citep{Hansen2004,Ligier2016,Cartwright2025} and numerical models \citep{Mergny2025cryst} indicate that although amorphous ice may be widespread at the immediate surface, the ice becomes predominantly crystalline at depths greater than a millimeter. The thermal skin depths associated with the thermal inertias reported in Section~\ref{ssec:results_ti} are typically on the order of $\sim$8~mm to 7~cm, and therefore exceed this thickness, implying that the ice probed in this study is expected to be mostly crystalline. We therefore restrict our analysis to the crystalline case.

We also assume that the temperature dependance in Equation~\ref{eq:I} can be neglected. Spectroscopic studies \citep{Hansen2004,Ligier2016,Cartwright2025} also suggest that grain sizes at the near-surface are typically of hundreds of micrometers. For such grain sizes, the temperature dependence of thermal inertia is weak \citep[see Figure 5 of][]{Ferrari2016} and can be neglected. Accordingly, we will assume constant thermophysical properties in Eq.~\ref{eq:I}, with $\rho_{\rm{ice}}$~=~918 kg m$^{-3}$ and $c_p$~=~805 J~kg$^{-1}$~K$^{-1}$, computed from \cite{Shulman2004} at a representative temperature of 90~K.

In this framework, thermal inertia depends on three parameters: grain size, porosity, and the quality of inter-grain contact \citep[see a full description in][]{Ferrari2016}. For the latter, we consider two end-member cases. In the "tight" model, grain contacts are assumed efficient, and the contact conductivity is computed following \citet{Johnson1971} and \citet{Gusarov2003}. In the "loose" model, grain contacts are assumed less efficient, with conductivity computed following \citet{Watson1964}. In both cases, the radiative conductivity is calculated following \cite{Breitbach1980} (assuming a temperature of~90~K) and added to the solid-state conductive term to obtain $k_{\rm eff,ice}$. Alternative formulations for conductive and radiative contributions are reviewed in \citet{Ferrari2016}. However, adopting different parameterizations modifies the inferred porosities derived in the following by less than $\sim$~10\% and does not affect the conclusions of this study.

For both the loose and tight models, we computed the resulting thermal inertia for grain sizes ranging from 1~$\mu$m to 10~cm. Although the upper bound exceeds the maximum grain sizes inferred from spectroscopic studies \citep[$\sim$ millimeter,][]{Hansen2004,Ligier2016,Cartwright2025}, comparing the derived grain sizes with spectroscopic constraints allows us to determine whether a given model requires unrealistically large grains and should therefore be ruled out.  Following \cite{Ferrari2016}, we considered porosities between 0.26 and 0.84 (Figure~\ref{fig:thermalinertia_theo_allice}). For each location and corresponding thermal inertia derived in  Section~\ref{ssec:constrainAlbTI}, we then determined the minimum and maximum grain sizes, as well as the minimum and maximum porosities, consistent with the derived thermal inertia (Figures~\ref{fig:illustrationmethod}c and d).

\subsection{Limits of the approach}
\subsubsection{Absorption of solar radiation within the ice \label{ssec:solarpen}} Our model assumes that solar energy is entirely absorbed within the  uppermost layer of the surface and is not distributed vertically within the ice. However, \cite{Brown1987} and \cite{Matson1989}  suggested that partial absorption of solar radiation at depth within  the ice could significantly heat subsurface layers, producing a solid-state greenhouse effect capable of modifying both surface and  subsurface temperatures. This effect may be particularly important for low thermal inertia surfaces such as those of icy satellites \citep{Ferrari2018}. 

Because water ice is opaque in the thermal infrared, and because the surface is expected to be composed of small water ice grains \citep[typically hundreds of microns according to spectroscopic studies, e.g.,][]{Ligier2016}, which limit radiative heat transport, heat generated by the absorption of solar radiation at the surface should be transported at depth by conduction. On icy satellite surfaces, where thermal conductivity is low \citep{Ferrari2018} the heat transport is not efficient. As a result, the surface warms more slowly in the morning than in the case where all solar energy is deposited at the surface (e.g., Figure~\ref{fig:illustration_solarabsinfround}), and the temperature maximum occurs later in the afternoon.  After the peak is reached, surface cooling can be relatively rapid, as incoming solar flux decreases while subsurface heat has already been conductively redistributed. This morning–afternoon asymmetry is a  characteristic signature of volumetric absorption of solar radiation.

The limited PPR coverage around the diurnal temperature maximum prevents us from directly detecting such an asymmetry and therefore from constraining this effect observationally. 
Nevertheless, using the large-scale binned PPR dataset of \cite{Rathbun2010} (see, for instance, their Figures 2a and 2c), which exhibits a relatively symmetric temperature curve around local noon, and implementing subsurface energy deposition in KRC following the approach of \cite{Urquhart1996}, we find that any solar absorption, if present, is likely confined to the uppermost centimeter of the surface. This result is consistent with the model of \cite{Urquhart1996}, who, based on Voyager observations, inferred a solar penetration depth smaller than 2.2~cm.

Both our model and that of \cite{Urquhart1996} operate in broadband and are therefore not spectrally resolved in the visible and infrared.  However, \cite{Brandt1993} showed that broadband treatments may overestimate the impact of solar absorption on surface and subsurface temperatures on Europa due to the strong wavelength dependence of ice optical properties. Furthermore, \cite{Brandt1993} demonstrated (see their Figure~5) that for  grain sizes of $\sim$50-100~$\mu$m, typical of Europa \citep{Hansen2004,Ligier2016,Cartwright2025}, most solar energy should be absorbed within the first few millimeters of the surface, well above the thermal skin depth probed here (up to $\sim$~6~cm). We therefore conclude that neglecting explicit volumetric absorption of solar radiation is unlikely to bias our results or the derived thermophysical parameters.  We acknowledge, however, that this reasoning strictly applies to relatively pure water ice. The presence of contaminants, such as carbonaceous material and Mg- and Na-sulfate hydrates \citep{Durham2010,McCord2001,Durham2010,Ligier2016,King2022}, which are strongly absorbing at visible wavelengths, could enhance near-surface energy deposition and potentially alter the thermal response. In addition, such contaminants may reduce the penetration depth of solar radiation within the ice through a decrease in surface albedo, further modifying the vertical distribution of absorbed energy. This effect is expected to be most significant in equatorial trailing hemisphere regions, where the abundance of non-ice material is highest \citep{Durham2010,McCord2001,Durham2010,Ligier2016,King2022}. Future observations with improved spatial coverage and a more complete sampling of the diurnal cycle in these regions will be necessary to assess whether such effects measurably influence the inferred thermophysical properties.

\subsubsection{Lateral heterogeneities and sub-pixel mixing \label{ssec:mixing}\label{ssec:subpixmixing}}
Given the low spatial resolution of a single PPR measurement \citep[typically $\sim$100~km,][]{Rathbun2010} before processing, the  radiance measured by the instrument can be affected by lateral heterogeneities within the instrument field of view, induced by variations in topography, surface roughness, albedo (e.g., bright ice versus dark, non-ice materials), or thermal inertia (e.g., porous ice versus coherent ice blocks). High-resolution mapping of albedo  \citep{Mergny2025} and surface roughness \citep{Steinbrgge2020} have shown a large variability of these properties at local scales. It is therefore unlikely that a single PPR measurement reflects a uniform surface unit, but rather that it samples a mixture of such heterogeneities.

Several techniques can be used to investigate this effect. One approach is to diagnose the mixing of materials with different thermophysical properties based on the apparent anisothermality between observations at two thermal infrared wavelengths, as developed and validated for Mars \citep{Christensen1986,Nowicki2007,McKeeby2022} and the Moon \citep{Bandfield2011}. This technique will notably be used with future measurements from E-THEMIS (Europa Thermal Emission Imaging System) onboard Europa Clipper \citep{Christensen2024}. However, it cannot be applied here because we only have a single brightness temperature at a given PPR wavelength \citep[the latter depending on the filter used for the observation, see][]{Russell1992}.

Another possibility is to model the measured surface temperature by assuming a mixture of different terrains and then derive the fraction of each terrain by fitting the modeled mixture temperatures to the observations, as performed on Mars by \cite{PUTZIG2007a,PUTZIG2007b}. However, given the large uncertainties regarding the diversity of materials across Europa's surface and their associated albedo and thermal inertia, together with the limited number of measurements, the problem is poorly constrained and cannot be robustly addressed here.  We illustrate this limitation in Appendix~\ref{app:mixing} by applying the method to a simple two-terrain mixing model. In particular, we find that a given retrieved pair of albedo and thermal inertia  is compatible with an infinite family of sub-pixel configurations and cannot, on its own, distinguish a homogeneous surface from a mixture. Hence, we acknowledge that the surface properties derived in this study may be biased by lateral heterogeneities and should be interpreted as apparent surface albedo and thermal inertia. Future observations at higher resolution by JUICE and Europa Clipper will help mitigate this issue. Vertical heterogeneities are discussed in Section~\ref{ssec:layering}.

In addition, at high latitudes, PPR measurements are generally acquired at high emission angles (typically $\sim$45\textdegree\ at 45\textdegree\ latitude, versus $\sim$0\textdegree\ at the equator). Combined with the large solar phase angles of the observations (on average $\sim$80\textdegree), this geometry increases the sensitivity to cold, shadowed facets within the field of view at high latitudes, whereas near-equatorial observations acquired close to zero emission angle  preferentially sample the warmer, directly illuminated surfaces. This geometric effect could bias our estimation of the thermal inertia,  as it would tend to artificially lower the derived thermal inertia at  high latitudes relative to equatorial regions. Using KRC, we modeled the brightness temperatures measured by PPR for  various slope distributions observed at different emission angles. Our results suggest that this effect is on the order of 1-2~K at typical PPR emission angles, remaining within the 2~K uncertainty of the measurements.

\section{Results \label{sec:Results}}
\begin{figure*}[h!] 
    \centering \includegraphics[width=\textwidth]{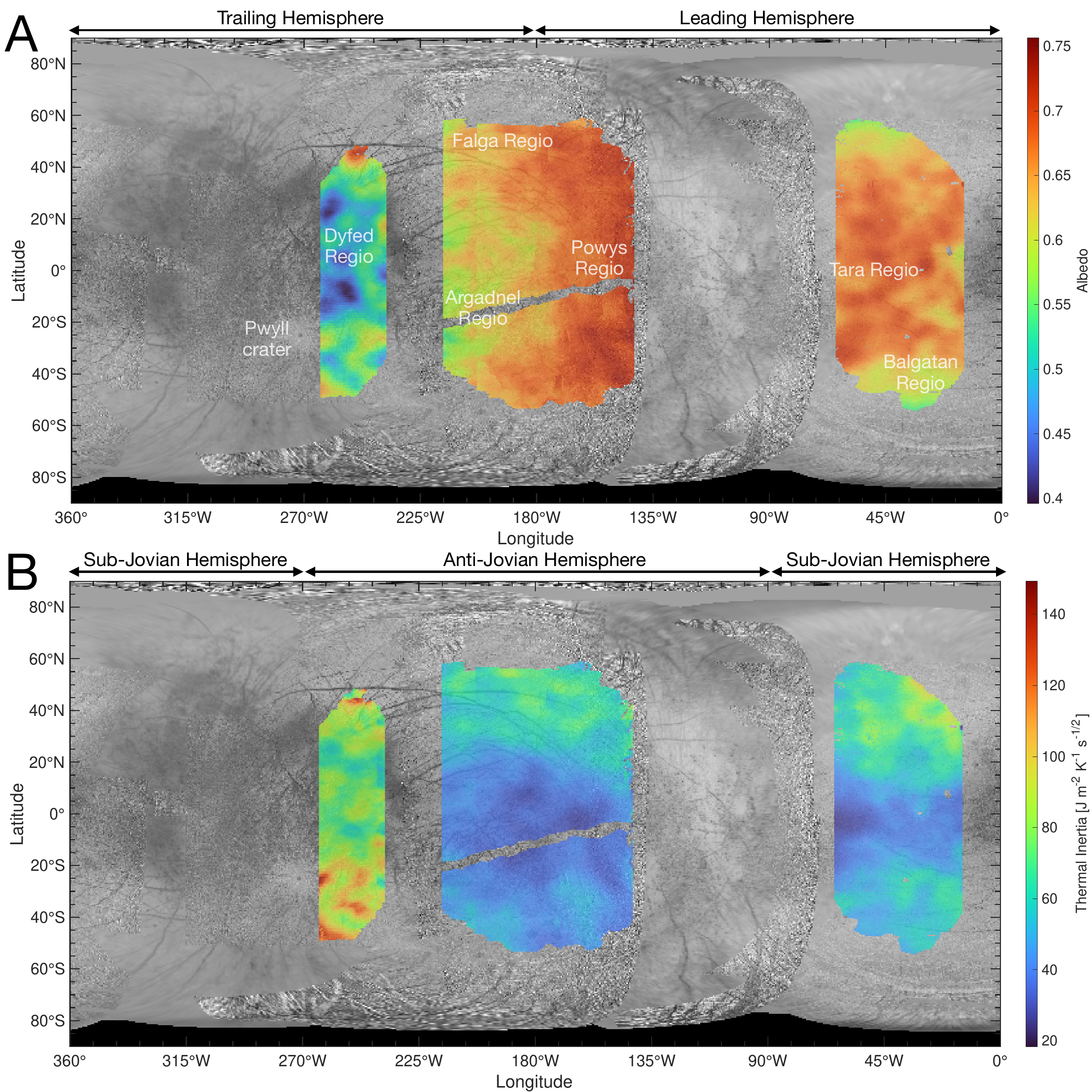} \caption{Bond albedo (a) and thermal inertia (b) maps derived from PPR measurements in this study. Background is the USGS Europa Voyager-Galileo SSI Global Mosaic of Europa \citep{Becker2010}.} 
    \label{fig:albedo_timap}
    \end{figure*}

The albedo and thermal inertia maps derived from our analysis of PPR measurements are presented in Figure~\ref{fig:albedo_timap}. The mean root-mean-square error between surface temperatures modeled with KRC using these maps and the PPR measurements is 0.38~$\pm$~0.4~K (standard deviation at 1$\sigma$). Hence, all PPR measurements can be explained by passive solar heating alone and do not require the presence of an endogenic heat source, as proposed by \cite{Rathbun2010,Rathbun2014,Rathbun2020} and \cite{Howes2025}.

\subsection{Bond albedo}
We find albedo values ranging from 0.39 to 0.76, with a mean and standard deviation values of 0.64~$\pm$~0.06 (1$\sigma$). The albedo exhibits strong longitudinal variations. In the leading hemisphere, the average albedo is 0.67~$\pm$~0.03 (1$\sigma$) and remains relatively homogeneous with latitude. In contrast, the trailing hemisphere appears darker, with an average albedo of 0.59 and stronger regional variability (standard deviation of 0.07 at 1$\sigma$). The albedo decreases from 150\textdegree~W toward the apex of the trailing hemisphere, where it reaches values as low as $\le$0.45. Latitudinal variations are also significant, with some equatorial regions exhibiting albedo values as low as 0.45, while brighter terrains are observed at higher latitudes and, locally, in the vicinity of the Pwyll crater. The low albedo in the equatorial regions of the trailing hemisphere should be mainly caused by continuous radiolytic alteration, which has made it darker and redder than the leading hemisphere \citep{Carlson2009}.

The albedo values derived here are overall consistent with those reported by \cite{Rathbun2010}. While there is good agreement between our albedo map and that of \cite{Rathbun2010} in the bright equatorial regions of the leading hemisphere, differences are observed in the darker regions of the trailing hemisphere, where our albedo values are higher by approximately 0.05–0.1. Overall, our albedo remains relatively constant with latitude, whereas \cite{Rathbun2010} reported higher values at high latitudes compared to equatorial regions. However, their study did not adjust solar illumination as a function of latitude, whereas we explicitly account for this effect, which likely explains the discrepancy. Our albedo values (0.39–0.76) are also consistent with those from the reanalysis of PPR data by \cite{Howes2025} (0.375–0.75), providing mutual cross-validation of the two independent analyses despite the use of different thermal models and approaches.

The derived albedo values are also consistent with visual patterns observed in the USGS Europa Voyager-Galileo SSI Global Mosaic of Europa \citep{Becker2010}: darker areas correspond to terrains with low albedo in our map, whereas bright regions are associated with higher albedo. \cite{Mergny2025} generated an albedo map combining the Global USGS Mosaic of Europa \citep{Becker2010} and Voyager-New Horizons photometric data. Comparison between our albedo map and theirs shows a pattern similar to that found with \cite{Rathbun2010}: there is good agreement in the bright areas of the leading hemisphere, while our albedo values are higher in the darker regions of Europa (mean difference of $\sim$0.08).

\subsection{Thermal inertia\label{ssec:results_ti}}
Thermal inertia derived from PPR measurements (Figure~\ref{fig:albedo_timap}b) ranges from 18 to 150~\tiu, with a mean and standard deviation values of 56~$\pm$~17 (1$\sigma$). Such low values are consistent with a particulate, uncompacted ice rather than solid water ice. These values aligned with those reported by \cite{Spencer1999} and \cite{Rathbun2010}, who found thermal inertia values between 40 and 150~\tiu. Some differences are observed with the results of \cite{Rathbun2010} at high latitudes, most likely due to differences in the treatment of solar illumination, leading to differences in the derived albedo and thermal inertia. Our results are broadly consistent with the independent PPR reanalysis of \cite{Howes2025}, who  report thermal inertia values ranging from 20 to 110 \tiu.  The range of values found here is also consistent with ALMA observations at 233~GHz \citep{Trumbo2018,Thelen2024}, although geographic discrepancies exist, as discussed below. Yet, the cold temperature residuals reported by \cite{Thelen2024} (see for instance their Figure 5) spatially correlate with our high thermal inertia regions, while their warm residuals coincide with our low thermal inertia zones, further supporting the consistency of our results and their spatial distribution.

\begin{figure}[h!] 
\centering
\includegraphics[width=0.45\textwidth]{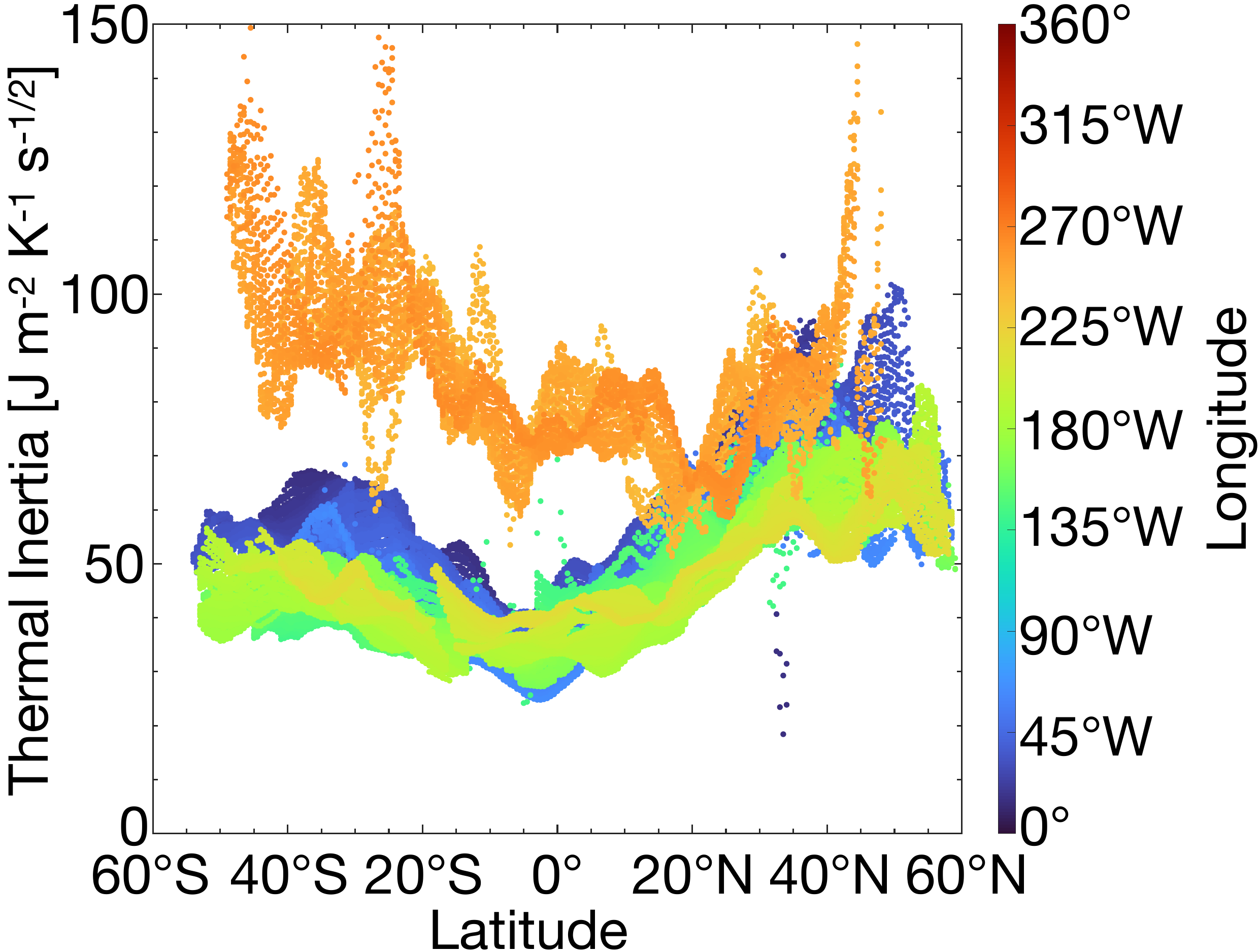} 
\caption{Thermal inertia of the surface as a function of latitude. Colors indicate west longitude. As a reminder, the leading hemisphere spans from 0\textdegree~to 180\textdegree W, and the trailing hemisphere from 180\textdegree W to 360\textdegree.}
\label{fig:thermalinertiavslatitude}
\end{figure}

Significant longitudinal variations are observed (a detailed discussion of correlations with geology is provided in Section~\ref{ssec:geology}). Thermal inertia in the leading hemisphere, notably at equatorial latitude, is lower than in the trailing hemisphere: on average 48~$\pm$~12~\tiu~in the leading hemisphere versus 63~$\pm$~17~\tiu~(1$\sigma$) in the trailing hemisphere in equatorial regions. This region of higher thermal inertia between 220\textdegree~W and 250\textdegree~W correlates with near-infrared spectroscopic studies suggesting the presence of non-ice components at the surface \citep[e.g.,][]{Ligier2016}, such as hydrated sulfuric materials, which may increase the thermal inertia by acting as a cementing agent within the porous ice  (see a more complete discussion in Section~\ref{ssec:traillead}). In the southern hemisphere, thermal inertia is higher in the trailing hemisphere than in the northern hemisphere, mainly due to the presence of high thermal inertia ejecta from the Pwyll crater (see also Section~\ref{ssec:geology}). At northern latitudes, except for a few locations, thermal inertia values are generally similar between the trailing and leading hemispheres (see also Figure~\ref{fig:thermalinertiavslatitude}).

Significant latitudinal variations are also observed (Figure~\ref{fig:thermalinertiavslatitude}), particularly on the leading hemisphere. There (blue and green dots in Figure~\ref{fig:thermalinertiavslatitude}), thermal inertia appears to decrease toward the equator: the median value between $\pm$~15\textdegree~latitude is 39~$\pm$~7~\tiu~(1$\sigma$), compared to 63~$\pm$~7~\tiu~(1$\sigma$) at higher latitude in the northern hemisphere and 47~$\pm$~8~\tiu~(1$\sigma$) in the southern hemisphere. Significant latitudinal variations are also observed (Figure~\ref{fig:thermalinertiavslatitude}), particularly on the leading hemisphere. A similar trend is present on the trailing hemisphere (orange dots). Note that on the leading hemisphere, for a given absolute latitude, thermal inertia is higher in the northern hemisphere than in the southern hemisphere. No firm conclusion can be drawn for the trailing hemisphere due to the uneven coverage of high latitudes between north and south. These variations cannot be explained by albedo effects, as albedo remains relatively constant within each longitude band. Also, when using the albedo values from \cite{Mergny2025}, we still recover a minimum thermal inertia at the equator. Possible explanations are discussed in Section~\ref{ssec:sinteringlat}. 

These conclusions are consistent with the interpretation of \cite{Spencer1999}, based on preliminary analyses of the first PPR data, and with the thermal inertia maps of \cite{Rathbun2010}, which also suggested a latitudinal trend. However, the absence of solar flux variations with latitude in the model of \cite{Rathbun2010} prevented definitive conclusions, a gap that our results now address. \cite{Howes2025} also found similar trend of lower thermal inertia near the equator of the leading hemisphere, although the values they retrieved are slightly lower at mid and high latitudes compared to our results (by $\sim$20\%). Such difference might be linked to the different approach we use to retrieve the thermal inertia. Interestingly, the latitudinal trend we highlight in this study does not appear in the analysis of \cite{Trumbo2018} using ALMA data (see, for instance, their Figure~3). Several methodological differences may explain this discrepancy. In the model of \cite{Trumbo2018}, emissivity is allowed to vary whereas we fix it at 0.9, and albedo values are imposed through bolometric albedos extrapolated from Voyager measurements. Furthermore, the observations of \cite{Trumbo2018} probe depths of approximately 1.5–3~cm, whereas our diurnal cycle is sensitive to thermal depths ranging from about 3~mm to 6~cm. Consistent with this depth dependence, \cite{Thelen2024} find a higher thermal inertia at mid-latitudes in the leading hemisphere within the first centimeter, in agreement with our results, but report that latitudinal variations diminish at greater depths (see their Figure~5). Together, these findings suggest that thermal inertia becomes more uniform with depth, with the heterogeneity being concentrated in the near-surface layers investigated here.

\subsection{Grain size and porosity\label{sec:resultsgrainsize}}

We then retrieved the minimum and maximum grain sizes, as well as the minimum porosity of the surface that can explain these observations using the model described in Section~\ref{ssec:derivingrandp}. When using the loose model (Figure~\ref{fig:illustrationmethod}d), we find that the minimum grain size required to explain the measured thermal inertia is on average 4.7~$\pm$~3~mm (1$\sigma$; throughout, the 1$\sigma$ values quoted refer to the spatial variability of the retrieved properties across the map, computed from the nominal thermal inertia values without accounting for their uncertainty). This range of grain size is inconsistent with spectroscopic studies suggesting grain sizes between 10~$\mu$m and $\sim$1~mm \citep{Hansen2004,Ligier2016,CruzMermy_Selectionchemicalspecies_I2022,Cartwright2025, CruzMermy_MicrophysicsEuropaSurface_I2025}. A word of cautious here : spectroscopic observations are sensitive to the uppermost layer of the surface, typically sampling depths from a few microns to at most a few hundred microns, whereas the thermal inertia derived here reflects properties integrated over the thermal skin depth, which extends to several centimeters. One may assume that larger grains could be thus present at depth. Nevertheless, millimeter-scale grain sizes remain unlikely. Such grain sizes are more characteristic of compacted, glacier-like ice \citep{Warren2019}, and it is improbable that such dense, coarse-grained material would be present within only a few centimeters of Europa's surface given the combination of low gravity and active surface processes (see a more complete discussion in Section~\ref{ssec:surfaceprocesses}. In contrast, when using the tight model, the minimum required grain size is on average 2.1~$\pm$~3~$\mu$m (1$\sigma$), and the maximum grain size is 2.8~$\pm$~2~cm (1$\sigma$).  To assess the impact of the thermal inertia uncertainty itself, we propagated the 3$\sigma$ uncertainty on the thermal inertia derived in Section~\ref{ssec:constrainAlbTI} through the conductivity model with a Monte Carlo scheme (1000 Gaussian draws per pixel). For the tight model, this yields 95\% confidence intervals of 1~-~11~$\mu$m for the minimum grain size and 0.7~-~8.6~cm for the maximum grain size. The lower bound of 1~$\mu$m is not constrained by the data but set by the minimum grain size assumed in our model (see Section~\ref{ssec:derivingrandp}). While the maximum grain size is somewhat higher than the expected range, this model remains broadly consistent with the values derived from spectroscopic studies. We therefore conclude that the tight model with relatively small grain size better explains the observations, suggesting efficient thermal contacts between the ice grains at the surface. 

\begin{figure*}[h!] 
\centering
\includegraphics[width=\textwidth]{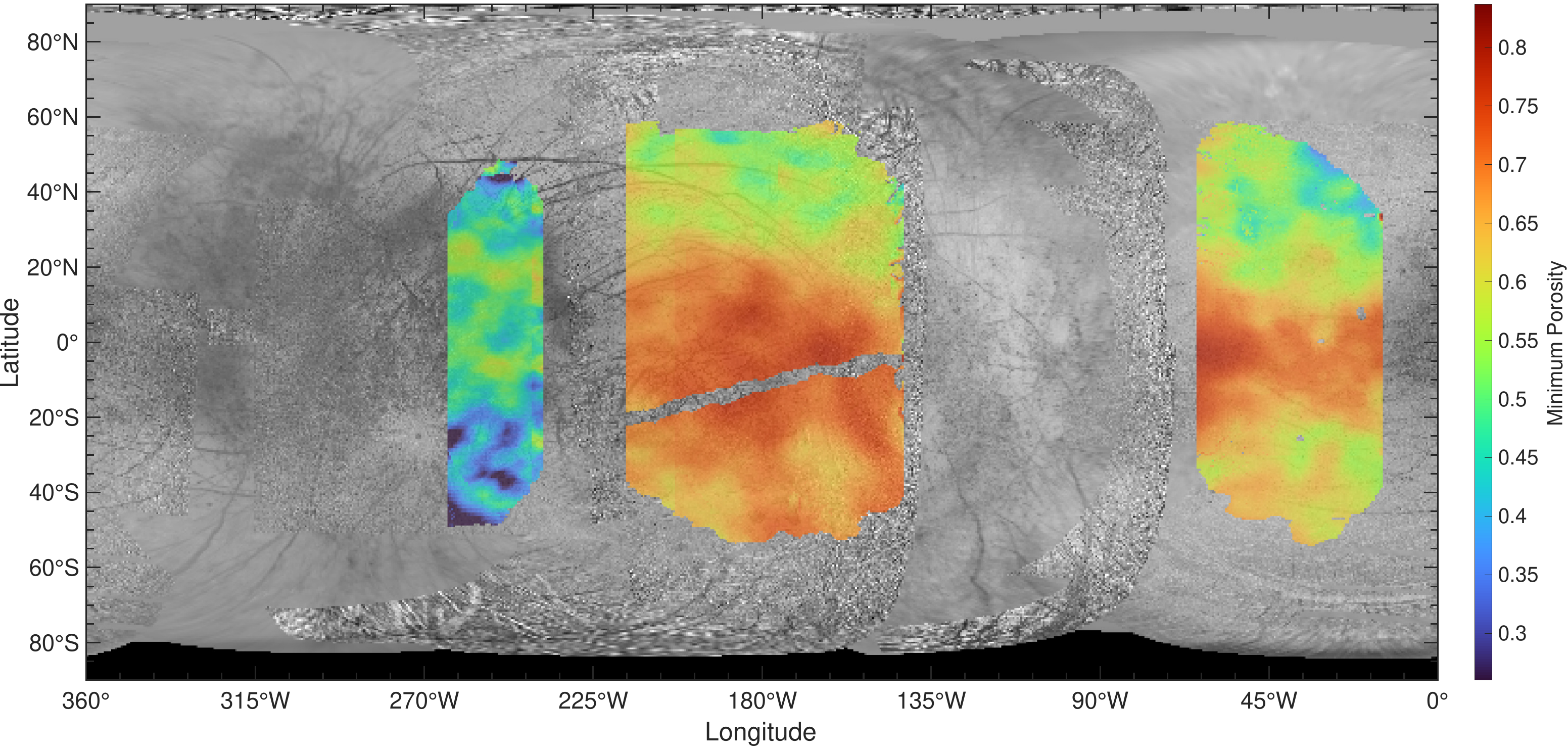} 
\caption{Minimum porosity required to explain the thermal inertia shown in Figure~\ref{fig:albedo_timap}b according to the crystalline tight model described in Section~\ref{ssec:derivingrandp}. Background is the USGS Europa Voyager-Galileo SSI Global Mosaic of Europa \citep{Becker2010}.}
\label{fig:porositymap}
\end{figure*}

Porosity values derived using this model are presented in Figure~\ref{fig:porositymap}. Note that we extract the minimum porosity over the full grain-size range of 1~$\mu$m~–~10~cm. Restricting the range to 1~$\mu$m~–~1~mm increases the minimum porosity by approximately 5\%. The average porosity in Figure is 0.61~$\pm$~0.1 (1$\sigma$), consistent with the porosity of the upper centimeter of Europa's surface derived from ALMA observations \citep[0.64~$\pm$~0.08,][]{Thelen2024}. Propagating the thermal inertia uncertainty as above yields a 95\% confidence interval of 0.36~-~0.76 on the minimum porosity.  The spatial distribution of porosity follows that of thermal inertia: maximum porosities (generally higher than 0.7) are found in equatorial regions of the leading hemisphere, whereas the trailing hemisphere and higher latitudes exhibit lower porosities (between 0.45 and 0.6). For some locations (e.g., Pwyll crater), the thermal inertia is too high to constrain the porosity: the entire assumed porosity range (0.26-0.84) in our conductivity models produces acceptable fits, leading to a degenerate solution. In these cases, only a lower bound of 0.26 can be reported, corresponding to the minimum value assumed in our model. The elevated thermal inertia at Pwyll crater also argues against the presence of amorphous ice at these locations, which is typically associated with low thermal inertia material \citep[Figure~\ref{fig:illustrationmethod_amorphous},][]{Ferrari2016}.

If amorphous ice is assumed, no constraint can be placed on the quality of grain contacts. Indeed, both the loose and tight models (Figure~\ref{fig:illustrationmethod_amorphous}) yield minimum and maximum grain sizes consistent with spectroscopic estimates. For the tight model, an upper limit on porosity of $\sim$0.5 can be inferred. For the loose model, no meaningful constraint on porosity is obtained, although the grain size can be constrained between 180 and 800~$\mu$m. Spectroscopic studies suggest that amorphous ice should be confined to the first millimeter of the surface \citep{Hansen2004,Ligier2016,Cartwright2025} at $\sim100~\mu$m grain size \citep{CruzMermy_MicrophysicsEuropaSurface_I2025}, with crystalline ice dominant at greater depths. Even in equatorial regions with the lowest thermal inertia, the thermal skin depth exceeds 3~mm, implying that most of the thermally probed ice is crystalline. We therefore consider the porosity derived from the crystalline tight model to be more reliable and present the amorphous case only for completeness.

\section{Discussion \label{sec:Discussion}}
\subsection{Correlation with Europa's geology \label{ssec:geology}}

\begin{figure}
\centering
\includegraphics[width=\linewidth]{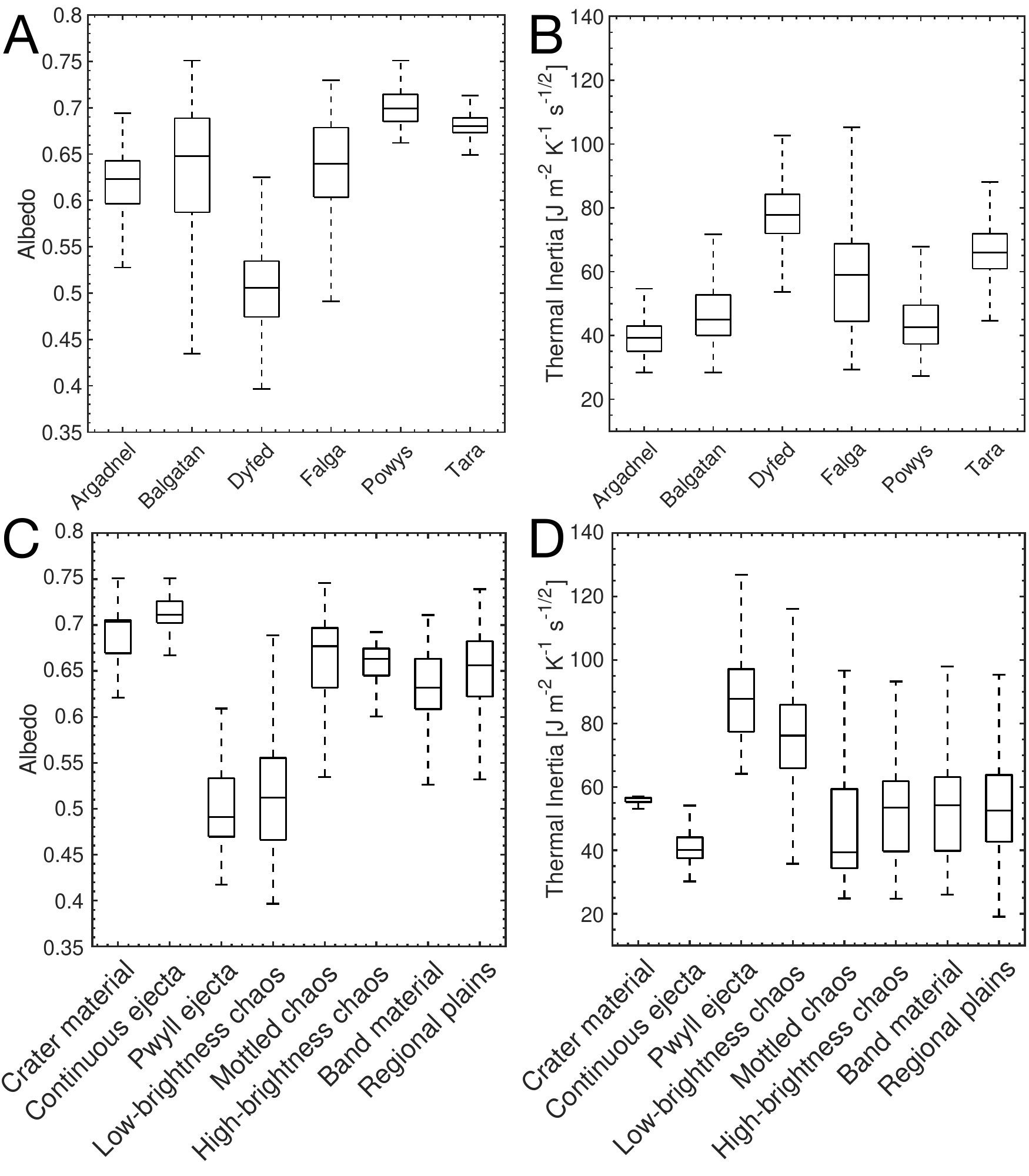}
\caption{Box-and-whisker diagrams of the albedo (a, c) and thermal inertia (b, d) of Europa's surface. Panels (a) and (b) show distributions extracted from physiographic regions defined by the IAU, while panels (c) and (d) correspond to geomorphological units defined by the geological map of Europa of \cite{Leonard2024}. For each box, the central line indicates the median, box edges denote the 25th and 75th percentiles, and whiskers extend to the most extreme non-outlier values. }
\label{fig:geologicalunit}
\end{figure}

On Mars, surface thermophysical properties are well correlated with geology, as both albedo and thermal inertia have been shown to reflect the distribution of geomorphological units \citep[e.g.,][]{Putzig2005,Piqueux2024}. On the Moon, thermal inertia reliably traces impact structures but does not follow major geological units \citep{Hayne2017}. Here, we examine possible correlations between the retrieved surface properties (Figure~\ref{fig:albedo_timap}) and geological units. Figure~\ref{fig:geologicalunit} presents the distributions of albedo and thermal inertia extracted from (1) regions defined by the International Astronomical Union (IAU) and (2) geomorphological units from the geological map of \cite{Leonard2024}.

Considering the regions, variations in albedo and thermal inertia largely reflect the latitude and longitude dependencies described in Section~\ref{sec:Results} and illustrated in Figure~\ref{fig:albedo_timap}. Albedo is relatively uniform and high across most regions (Figure~\ref{fig:geologicalunit}a), with the exception of Dyfed, whose observations are concentrated over the darker trailing hemisphere. Similarly, for thermal inertia (Figure~\ref{fig:geologicalunit}b), the Dyfed region exhibits the highest values (median thermal inertia of 78~$\pm$~10~\tiu~(1$\sigma$)), while most other regions show median values between 40 and 50~\tiu. Falga presents a somewhat elevated thermal inertia (median thermal inertia of 59~$\pm$~16~\tiu~(1$\sigma$)), consistent with its central latitude of $\sim$30$\textdegree$N, where thermal inertia is enhanced relative to equatorial regions.

Turning to the geological units, albedo (Figure~\ref{fig:geologicalunit}c) shows little variation across most units, with the notable exceptions of low-albedo chaotic terrain and the high-albedo Pwyll crater ejecta. Regarding thermal inertia (Figure~\ref{fig:geologicalunit}d), crater materials and ejecta deposits show contrasting behaviors: the continuous ejecta of Pwyll exhibits a significantly elevated thermal inertia (median thermal inertia of 88~$\pm$~16~\tiu~(1$\sigma$)), consistent with the emplacement of more compacted material during the impact event, as previously reported by \cite{Spencer1999}.  High inertia values in the ejecta blankets of  fresh impact craters are likewise observed on icy moons such as Rhea, with Inktomi standing out as a clear example \cite{Howett2014,Bonnefoy2020}.  In contrast, other crater-related units show median thermal inertia values between 40 and 56~\tiu, comparable to the background surface, including the Taliesin crater region. \cite{Thelen2024} reached a similar conclusion using ALMA data. This likely reflects that older impact crater surfaces are processed (e.g., by space weathering) to depths comparable or larger to those sensed by infrared/millimetric wavelengths, such that they no longer exhibit a distinct inertia signature. Chaotic terrains display a bimodal behavior, with low-brightness chaos presenting elevated thermal inertia (median thermal inertia of 76~$\pm$~18~\tiu~(1$\sigma$)) and high-brightness chaos showing lower values (median thermal inertia of 53~$\pm$~14~\tiu~(1$\sigma$)). No clear systematic relationship between chaos type and thermal inertia is therefore identified, consistent with the findings of \cite{Rathbun2014}. Regional plains and band materials show relatively uniform thermophysical properties; the lack of any thermal inertia signature in band material likely reflects the fact that these narrow features remain unresolved given the spatial smoothing applied during the construction of the thermal dataset (Section~\ref{ssec:subpixmixing}). Overall, with the exception of the Pwyll ejecta, no correlation between thermal inertia and geological units is identified on Europa, consistent with previous studies by \cite{Rathbun2014}, \citet{Trumbo2017},  \cite{Thelen2024} and \cite{Howes2025}.

\subsection{Spatial distribution of thermal inertia and surface processes \label{ssec:surfaceprocesses}}

Surface properties on Europa, and in particular thermal inertia, show large spatial variations, with notably higher thermal inertia in the trailing hemisphere (Figure~\ref{fig:albedo_timap}) and lower thermal inertia in equatorial regions (Figure~\ref{fig:thermalinertiavslatitude}). The properties of the ice at the surface may be related to the evolution of its microphysical characteristics through sintering processes \citep[see a review in ][]{Blackford2007}, or through condensation–sublimation cycles of water ice \citep[a process called thermal segregation;][]{Spencer1987,Sorli2026}, as well as through  interactions with the external environment (so-called exogenic processes) or internal activity (endogenic processes). Regarding the latter, the previous section showing the absence of a clear correlation with geological units suggests that the surface thermophysical properties derived from the PPR data are not primarily controlled by endogenic processes. No definitive conclusion can however be drawn because (1) the spatial averaging inherent in the temperature map generation may have smoothed out thermal contrasts between distinct geological units (e.g., band materials), thereby diluting potential correlations, and (2) the initial spatial resolution of the data may have prevented the detection of homogeneous geological units \citep[e.g., plume ejecta that could exhibit different thermal inertia,][]{Hayne2025}). We also do not investigate thermal segregation process \citep{Spencer1987,Sorli2026} in the absence of a model describing vapor transport in Europa's exosphere. Also, gravity-driven compaction is not considered: given Europa's low surface gravity \citep[$\sim$~1.31~m~s$^{-2}$,]{Anderson1998}, this process is expected to be inefficient within the first centimeters of the subsurface probed in this study \citep{Mergny2024compaction}.

We therefore focus on the role of sintering processes and exogenic processes, particularly meteoritic gardening and sputtering, to explain the observed distribution of thermal inertia.

\subsubsection{Absence of the PacMan feature on Europa \label{ssec:pacman}}
The icy satellites of Saturn, Mimas, Tethys, and Dione, show evidence of an IR/UV color anomaly and anomalously high thermal inertia in equatorial regions on their leading hemispheres, known as the PacMan feature \citep{Howett2011,Schenk2011,Howett2012,Howett2014}. This feature is spatially correlated with areas preferentially bombarded by high-energy electrons \citep{Nordheim2017}: at these moons, electrons with energies above $\sim$~1~MeV, which penetrate to depths comparable to the thermal skin depth, precipitate in a lens-shaped pattern centered on the equatorial regions of the leading hemisphere. It has therefore been hypothesized that these high-energy electrons increase the contact area between grains, enhancing their thermal conductivity and hence their thermal inertia \citep{Howett2011,Schenk2011,Schaible2017}.

At Europa, \citet{Nordheim2018} showed that high-energy electrons (in the MeV range, penetrating to the thermal skin depth) should bombard the equatorial regions of both the leading and trailing hemispheres in lens-shaped patterns similar to those predicted for the Saturnian moons \citep{Paranicas2014,Nordheim2017}, but with a preference for the trailing hemisphere. One might therefore expect a PacMan-like anomaly at the equator, most pronounced on the trailing hemisphere. However, the opposite trend is observed here, with lower thermal inertia at the equator on both hemispheres (Figure~\ref{fig:albedo_timap}b, \ref{fig:thermalinertiavslatitude}), whereas higher values are observed on Mimas, Tethys, and Dione. We acknowledge that the equatorial regions of the trailing hemisphere exhibit a higher thermal inertia than those of the leading hemisphere; this difference, however, is unlikely to be caused by an electron-induced effect, but rather reflects a difference in composition and weathering processes, as explained in Section~\ref{ssec:traillead}. 

The presence of a PacMan anomaly on some of Saturn's icy moons, and its absence on Europa, may be explained by differences in the microscopic properties of the ice on these satellites. The electron-induced sintering mechanism, which could be responsible for such high thermal inertia as modeled by \citet{Schaible2017}, requires the initial icy regolith to be relatively compact (porosity lower than 0.65) and composed of small ($\leq$ 5~$\mu$m) grains or grain aggregates, and/or to have poor-quality contacts between grains (and therefore low contact conductivity). These conditions are met on Mimas, with typical grain sizes of $\leq$~20~$\mu$m \citep{Buratti2011} and poor contact between grains in low thermal inertia regions \citep{Ferrari2016}.  On Europa, however, these conditions are not all met. First, typical grain sizes generally range from tens to hundreds of micrometers \citep{Hansen2004,Ligier2016,Cartwright2025,CruzMermy_MicrophysicsEuropaSurface_I2025}. Second, the porosity condition (lower than $\sim$~0.65) is satisfied only on the trailing hemisphere (0.45~-~0.60, Figure~\ref{fig:porositymap}), not on the leading hemisphere (greater than 0.7). Finally, the relatively high thermal inertia measured on Europa compared to Mimas \citep[56~\tiu\ in this study versus 16~\tiu\ on Mimas,][]{Howett2011} is only consistent with a regolith characterized by good contact conductivity rather than loose grain contacts (Figures~\ref{fig:illustrationmethod}c, d; Section~\ref{sec:resultsgrainsize}). Hence, the conditions required to produce electron-induced sintering are not jointly met on Europa's surface, explaining the absence of a PacMan feature correlated with high-energy electron bombardment.

\subsubsection{Latitudinal variations in thermal inertia and the equatorial low-inertia region\label{ssec:sinteringlat}}
The thermal conductivity of icy materials can increase as a result of sintering. One of the main sintering processes studied in planetary science is isothermal sintering. Isothermal sintering in snow is the process by which ice grains, at constant temperature, progressively bond through migration of water vapor driven by gradients in saturated vapor pressure associated with local surface curvature, forming ice bridges that increase thermal conductivity and thus thermal inertia \citep[see a full review in][]{Blackford2007}. This process is highly sensitive to temperature because of the exponential dependence of saturated vapor pressure on temperature \citep{Murphy2005}.

The presence of a band of low thermal inertia at the equator is therefore surprising. Indeed, because these regions receive more insolation than higher latitudes, isothermal sintering \citep{Molaro2019,Mergny2024lunaicy} should, at equivalent albedo, be more efficient at low latitudes and should increase the thermal inertia there relative to higher latitudes, contrasting with the observations presented here (Figure~\ref{fig:albedo_timap}). \cite{Mergny2024lunaicy}, using the LunaIcy isothermal sintering model, show (their Figure~9) that for the albedo and porosity derived at the equator (Figures~\ref{fig:albedo_timap}, \ref{fig:porositymap}), the thermal conductivity of the ice should increase by a factor of 5 to 30 (corresponding to an increase in thermal inertia by a factor of 2 to 5) in only $\sim$~1~Myr \citep[versus $\sim$~30~Myr for the estimated surface age,][]{Pappalardo1998}.

Isothermal sintering may only occur when the pores within the ice are not connected and when no vertical vapor transport takes place within the ice or between the subsurface and the external environment. However, our results instead suggest that the regolith is highly porous (Figure~\ref{fig:porositymap}), implying that significant vapor transport between pores may occur.  \citet{Mergny2024lunaicy} and \cite{Mergny2026} noted that, in such porous structures, an alternative diffusion mechanism, known as temperature-gradient metamorphism on Earth \citep{Colbeck1983}, may happen if a thermal gradient exists withing the regolith.
This hypothesis is further supported by the low thermal inertia values derived in this study, favoring strong near-surface thermal gradients. Using the albedo and thermal inertia derived in Section~\ref{sec:Results}, we predict temperature gradients exceeding $\sim$~30~K within the upper 10~cm of the subsurface, consistent with numerical simulations by \cite{Mergny2024multiheat}. Such temperature gradients induce strong gradients in saturated vapor pressure, the latter varying from $\sim10^{-21}$~Pa at 80~K to $\sim10^{-15}$~Pa at 100~K and up to $\sim~10^{-8}$~Pa at 130~K, suggesting the presence of vapor fluxes within the upper centimeters of the subsurface.

In line with this, recent work from \citet{Mergny2026} proposed a formation scenario for the Galilean moons regolith involving temperature-gradient metamorphism. On Earth, such thermal gradients drive a process known as temperature-gradient metamorphism in snowpacks \citep{Colbeck1983}. Vertical thermal/vapor pressure gradients within the ice leads to the sublimation on the warm side of ice grains and re-deposition on their colder side. In contrast to isothermal sintering, this process can destroy ice bridges and promote the growth of large faceted crystals or depth hoar \citep{Giddings1962,Colbeck1983}. Terrestrial studies suggest that this mechanism becomes significant when the thermal gradient exceeds $\sim$10~K~m$^{-1}$ \citep{Colbeck1983,Marbouty1980}. The thermal gradients predicted for Europa are one order of magnitude larger. It is therefore plausible that similar processes occur there, leading to subsurface sintering rather than at the surface. This mechanism could explain the presence of a low thermal inertia surface layer in equatorial regions, overlying a less porous layer with higher thermal inertia at depth \citep[50\% porosity at depth of 10–20~cm;][]{Thelen2024}.

An important caveat, however, is that we cannot definitively conclude on the role of this process without dedicated modeling. On Earth, temperature-gradient sintering occurs at temperatures roughly twice as high as those on Europa and under much higher partial vapor pressures. It is therefore possible that, under Europa's extremely low vapor pressures, the process may in fact be negligible. Modeling this mechanism and comparing it with isothermal sintering and surface processes therefore represents an important avenue for improving our understanding of icy surfaces in the Solar System.

In addition to temperature-gradient metamorphism, the apparent inefficiency of isothermal sintering at low latitudes may also reflect a competition with surface erosion processes. Isothermal sintering is expected to operate primarily within the upper $\sim$~10~cm of the regolith over timescales of $\sim$1~Myr \citep[see Figure~8 of][]{Mergny2024lunaicy}, which coincides with the characteristic depth affected by impact gardening and sputtering erosion \citep[e.g., Figure~1 in][]{Chyba2001}. The high porosities inferred in this study may therefore reflect a competition between sintering and destructive processes such as sputtering and impact gardening, continuously disrupting grain contacts. In contrast, the lower porosities inferred at greater depths by \cite{Thelen2024} may indicate that these destructive processes become less efficient with depth, allowing sintering to progressively dominate and promote regolith compaction. At present, however, no coupled modeling has quantitatively assessed the relative kinetics of these competing processes under Europa conditions. Determining whether sintering is effectively suppressed by surface renewal or instead shifted to greater depths thus remains an important objective for future studies.

What could then explain the higher thermal inertia observed at mid-latitudes? \cite{Clark1983} suggested that regolith grain sizes, and therefore the thermal conductivity of ice on Europa, may be controlled by the sputtering rate. While sputtering can destroy small grains as stated previously, it may also promote the growth of larger grains, as sputtered water molecules may preferentially adsorb onto larger grains, potentially forming bonds between grains and thereby increasing thermal inertia \citep{Clark1983,Cassidy2013}. \cite{Cassidy2013} noted a significant correlation between the magnetospheric ion sputtering rate predicted by their model and the grain size of Europa's icy surface, suggesting an important role for sputtering. 

Updated magnetospheric ion fluxes from \cite{Nordheim2022} show that the flux of keV O$^{++}$ and S$^{+++}$ ions, thought to be the dominant ion sputtering agents at Europa \citep{Plainaki2010,Cassidy2013}, correlates well with the spatial distribution of thermal inertia presented in this study (compare their Figure~11 with our Figure~\ref{fig:albedo_timap}b, Figure~\ref{fig:appendix:magnetosphericflux}). Their results indicate that ion bombardment preferentially occurs at mid to high latitudes rather than at low latitudes, on the leading hemisphere and, to a lesser extent, the trailing hemisphere. To quantify this agreement (Figure~\ref{fig:appendix:magnetosphericflux}), we resampled \cite{Nordheim2022} modeled fluxes (their Figure~11) onto the grid of our retrieved thermal inertia map and computed the Pearson correlation coefficient over the common coverage, treating the leading and trailing hemispheres separately because they differ markedly in surface composition (Section~\ref{ssec:traillead}). Over the leading and anti-jovian hemispheres, thermal inertia and ion flux are strongly correlated (oxygen: $r = 0.67$; sulfur: $r = 0.67$), in line with the preferential bombardment of mid to high latitudes and the increase of thermal inertia away from the equator. The correlation is weaker over the trailing equatorial region (oxygen: $r = 0.47$; sulfur: $r = 0.46$), where the higher thermal inertia is more likely set by composition than by sputtering (Section~\ref{ssec:traillead}). These results suggest that ion sputtering can be a driver of sintering where the surface is comparatively clean water ice.

Finally, while we ruled out a significant role of electron-driven sintering at low latitudes in Section~\ref{ssec:pacman}, one might wonder whether this process becomes efficient at higher latitudes on the leading hemisphere, where the surface can be bombarded by high-energy ($>$50~MeV) electrons \citep{Nordheim2018}. However, this hypothesis appears unlikely for two reasons. First, the electron flux is actually lower at higher latitudes, reducing the possible occurrence of any electron-driven sintering. Second, grain sizes are larger at higher latitudes \citep[e.g.,][]{Ligier2016}, whereas the sintering mechanism driven by electron sputtering described by \cite{Schaible2017} requires small grains to be effective. We therefore conclude that electron-driven sintering is unlikely to play a significant role at any latitude on Europa's surface. However, as for temperature-gradient metamorphism, dedicated modeling and experimental studies are required to test and validate the contribution of magnetospheric ion and electron sputtering on the sintering of Europa's surface.

\subsection{Difference between the trailing and leading hemispheres \label{ssec:traillead}}

As noted in Section~\ref{ssec:results_ti}, the thermal inertia in the equatorial region of the trailing hemisphere is higher than in the leading hemisphere (median thermal inertia of 63~\tiu~vs.~48~\tiu). This asymmetry may partly reflect differences in surface processing between the two hemispheres. First, modeling by \cite{Nordheim2022} shows that the magnetospheric ion flux of keV O$^{++}$ and S$^{+++}$ ions is slightly higher at equatorial latitudes of the trailing hemisphere compared to the leading hemisphere (see their Figure~11). If sintering driven by magnetospheric ion sputtering is indeed the dominant process on Europa's surface, this difference in ion flux might partly explain the asymmetry in thermal inertia. Second, micrometeoritic bombardment, and hence impact gardening, is expected to be more important on the leading hemisphere \citep{Zahnle1998}, potentially contributing to lower thermal inertia there.

However, we propose that this asymmetry most likely results from differences in surface composition. As noted by \cite{Ligier2016} and \cite{King2022}, the trailing equatorial region is predominantly composed of sulfuric acid hydrates \citep[abundance of $\sim$~65\%;][]{Ligier2016,King2022}, whereas the leading equatorial region is primarily composed of water ice\citep[abundance of $\sim$~80\%;][]{Ligier2016}. Several studies have compared the thermal conductivity of salts with that of water ice \citep[see a review in][]{Durham2010}, generally concluding that hydrated salt materials have lower conductivity than pure ice. However, these studies typically consider compact slab ice rather than porous ice, the latter exhibiting much lower conductivity \citep{Ferrari2016}. Studies focusing on porous mixtures of ice and silicate minerals, representative of cometary materials, have shown that the presence of minerals can drastically increase thermal conductivity at low temperatures, by up to two to three orders of magnitude \citep{Seiferlin1996,Prialnik2004}. Similarly, \cite{Piqueux2009} concluded that the presence of salts within the porous regolith of Mars could increase the thermal conductivity of porous materials. We therefore hypothesize that a similar effect may occur on Europa, where sulfuric acid hydrates within the porous regolith could enhance the effective thermal conductivity and thereby increase the thermal inertia of the trailing hemisphere relative to the leading hemisphere. Dedicated modeling and laboratory experiments are required to test and quantify this effect.

\subsection{Vertical layering \label{ssec:layering}}
The thermophysical modeling performed in this study assumes homogeneous subsurface properties with depth. However, the subsurface may in reality be layered, affecting conduction in the near-subsurface. This limitation can be partially assessed by comparing our derived porosity profiles with those from previous studies. \cite{Domingue1997} and \cite{Hendrix2005}, using analyses of the solar phase curve obtained from visible and UV observations, showed that the upper micrometers of the surface have porosities of $\sim$0.8–0.95. \cite{Mishra2021}, revisiting the Galileo Near Infrared Mapping Spectrometer data, also found that the porosity of the upper micrometers should range between 0.81 and 0.97. Observations of Europa's rapid cooling during eclipse by \cite{Hansen1973} revealed that the upper millimeters of the surface have a very low thermal inertia of $\sim$14~\tiu, which can be associated with porosities higher than 0.8 assuming crystalline ice and tight contacts (Section \ref{ssec:derivingrandp}). Our results (Section~\ref{sec:resultsgrainsize}, Figure~\ref{fig:porositymap}) indicate that the first centimeters of the surface should have a minimum porosity of 0.61~$\pm$~0.1 (1$\sigma$). ALMA observations at 0.87~mm by \cite{Thelen2024} suggest that the porosity at 10–20~cm depth is 0.52~$\pm$~0.07. Since gravity on Europa is too weak to induce compaction in such shallow depths, this most likely results from 1) a balance between surface destruction processes and 2) sintering processes, as described in Section~\ref{ssec:sinteringlat}.

\begin{figure}
\centering
\includegraphics[width=\linewidth]{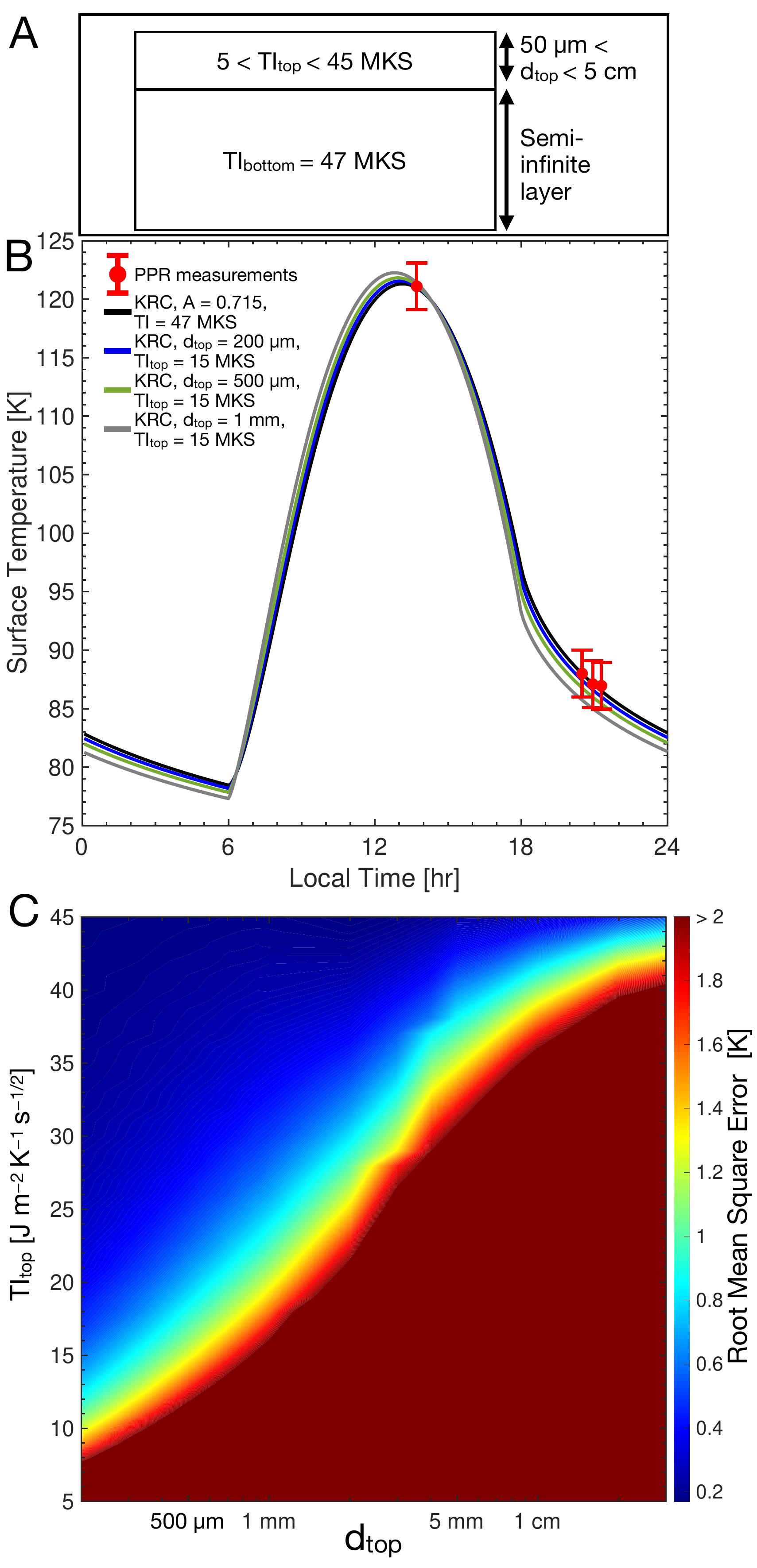}

\caption{Sensitivity of modeled surface temperatures to two-layer regolith properties. The thermal inertia unit is abbreviated as MKS in this figure. a) Schematic of the two-layer model used in this study. b) Surface temperatures modeled with KRC for several values of d$_{\rm{top}}$ at TI$_{\rm{top}}$ = 15~\tiu\ (colored curves), compared to the best-fit case (dark curve) and PPR measurements. c) Root mean square error as a function of d$_{\rm{top}}$ and TI$_{\rm{top}}$ between the modeled surface temperature and PPR measurements. Simulations are performed at the PPR measurement location (latitude 0.1\textdegree~N, 30\textdegree~W).}
\label{fig:layering}
\end{figure}

\cite{Mergny2024multiheat} showed (their Figure 4) that layering can significantly affect the diurnal temperature curve on Europa. In particular, nighttime cooling can be buffered in the presence of sharp layering, with an insulating material overlying a higher thermal inertia substrate. Given the limited number of data points over a full diurnal cycle measured by PPR at each location, it is difficult to determine whether such layering is present. The relative nighttime cooling observed at a few locations (see, e.g., Figure \ref{fig:illustrationmethod}a) does not support the presence of pronounced layering. The previous discussion also suggests gradual compaction with depth rather than abrupt layering, potentially mitigating this effect. Nevertheless, we quantitatively test this assumption using KRC and the PPR data. We consider a two-layer model (Figure \ref{fig:layering}a) with a variable top-layer thickness (d$_{\rm{top}}$ ranging from 50~$\mu$m to 5~cm) and thermal inertia (TI$_{\rm{top}}$ between 5 and 45~\tiu), overlying a semi-infinite bottom layer (TI$_{\rm{bottom}}$~=~47~\tiu). For each configuration, we compute the root mean square error between the PPR measurements and the KRC-modeled temperatures to identify the best-fitting parameters (Figures \ref{fig:layering}b,c).

The error analysis indicates that either a uniform model or an extremely thin surface layer is preferred (Figure \ref{fig:layering}c). If layering is present, particularly with a low thermal inertia layer overlying a higher thermal inertia substrate (TI$_{\rm{bottom}}$), the top layer must be extremely thin (preferably less than 1~mm) otherwise the misfit increases rapidly (Figure~\ref{fig:layering}b). If the low thermal inertia layer is interpreted as an amorphous ice layer (e.g., Figure~\ref{fig:illustrationmethod_amorphous}), then our results are consistent with spectroscopic studies suggesting that amorphous ice, potentially associated with low thermal inertia, is confined to the uppermost micrometers of the surface \citep{Hansen2004,Ligier2016,Cartwright2025}.

\subsection{Prediction of surface temperatures and applications to surface properties}

\begin{figure}
\centering
\includegraphics[width=\linewidth]{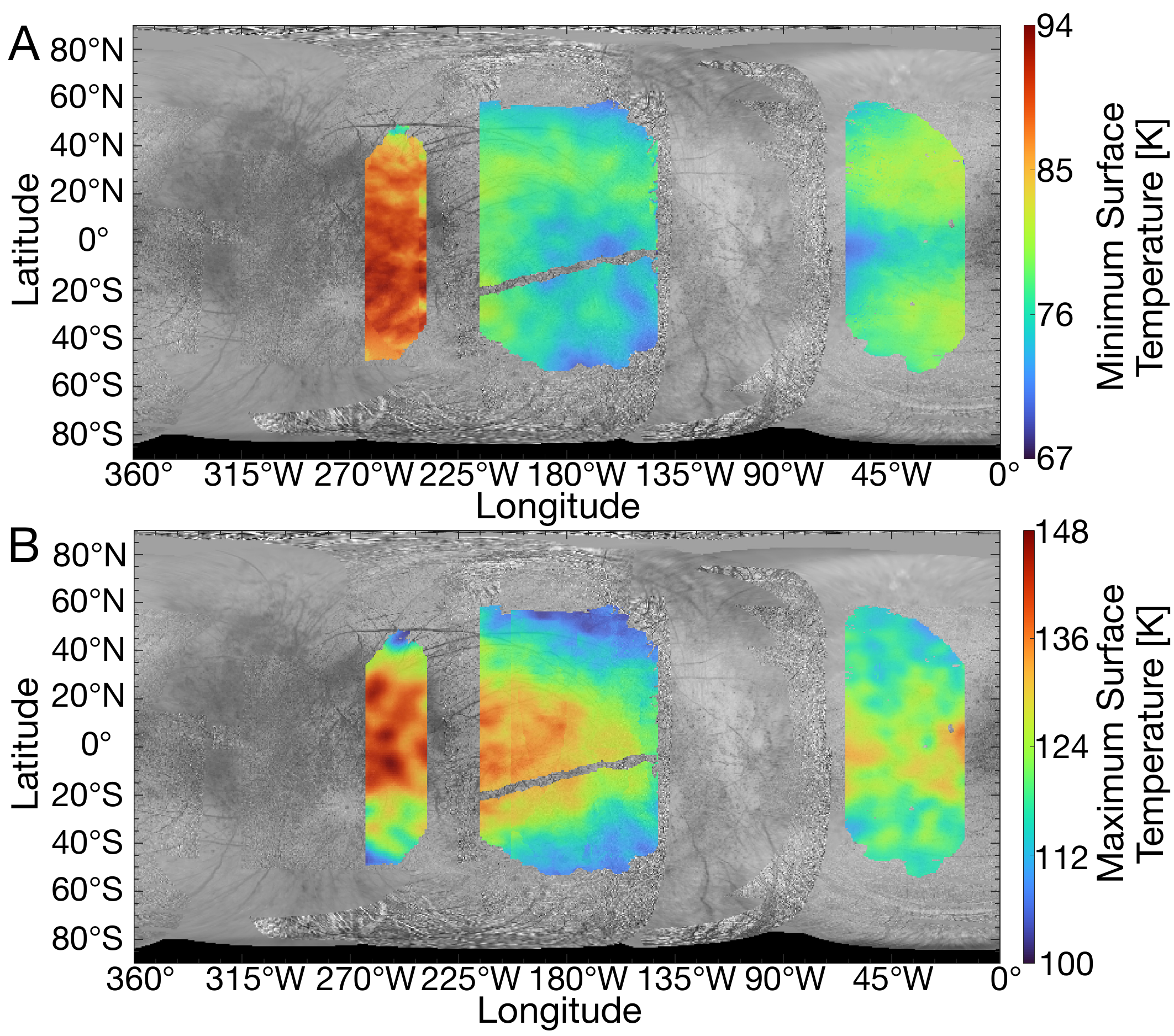}

\caption{Minimum (a) and maximum (b) surface temperatures over the jovian year simulated with KRC using the albedo and thermal inertia maps generated in Figure~\ref{fig:albedo_timap}. An emissivity of 0.9 is assumed. The range of temperatures predicted between 2030 and 2034, during the Europa Clipper mission, is shown in Figure~\ref{fig:TminTmax_ETHEM}.}
\label{fig:TminTmax}
\end{figure}

Using the thermal inertia and albedo maps generated in this study (Figure~\ref{fig:albedo_timap}), we use KRC to derive the range of possible surface temperatures at each location over a jovian year (Figure~\ref{fig:TminTmax}). Surface temperatures are expected to range between 67.5~K and 148~K, in agreement with previous estimates of surface temperatures on Europa \citep[e.g.,][]{Mergny2024multiheat}. Nighttime temperatures are expected to range between 67.5 and 93~K, being maximum in the trailing hemisphere as the latter has a higher thermal inertia, limiting nighttime cooling, and a higher albedo, increasing the amount of solar radiation absorbed at the surface. Similarly, the maximum daytime temperatures are expected to range between 101 and 148~K, and follow the distribution of the albedo, with maximum daytime temperatures predicted in the equatorial region of the trailing hemisphere of Europa exhibiting the lowest albedo. Maximum surface temperatures decrease with increasing latitude, as these high-latitude regions receive a lower amount of insolation compared to the equator.

\paragraph{Implications for ice properties:} The excursion to high surface temperatures during the day, notably for the low-latitude and low thermal inertia regions, has several implications regarding the ice properties. First, as noted by \cite{Mergny2024lunaicy}, high temperatures can drastically increase thermally driven sintering of the ice. Again, competing processes, as explained in Section~\ref{ssec:sinteringlat}, could mitigate the influence of this high-temperature excursion on the ice properties. Second, high daytime temperatures could promote the crystallization of water ice at the surface, counteracting radiolytic amorphization. Indeed, this process is temperature dependent, increasing exponentially with temperature \citep{Jenniskens1996}. In addition, the high porosity derived in these low thermal inertia regions should increase the crystallization rate, as the latter is also enhanced in high-porosity media \cite{Mitchell2017}. Although the final crystallization rate depends on the competition between thermally driven crystallization and radiation-driven amorphization \citep{Mergny2025cryst,Yoffe2026,Moingeon_reappraisalamorphizationkinetics_I2026}, maximum temperatures reached during the diurnal cycle should still promote the crystallization of ice in the shallow subsurface \citep[see for instance Figure~6 of][]{Yoffe2026}.

\paragraph{Implications for volatile/hydrated materials at the surface:} The high temperatures reached during daytime at low latitudes are too high to allow the cold trapping of volatiles such as NH$_3$, SO$_2$, or CO$_2$ \citep[see for instance Figure~1 of][]{Zhang2009}. As a result, volatile species such as CO$_2$, detected at the surface of Europa \citep[e.g.,][]{Trumbo2023,Villanueva_EndogenousCO2ice_S2023}, should be currently unstable and sublimating. Given the lower temperatures reached at higher latitudes due to reduced insolation, these regions might favor the preservation of volatile frost or deposits over geological timescales compared to the equator. Similarly, topography-induced cold trapping, as observed on Mercury \citep[e.g.,][]{Paige2013} or the Moon \citep[e.g.,][]{Paige2010,Schorghofer2021}, could preserve some reservoirs of volatiles. Such studies are left for future work.

Although the stability of hydrated minerals and acid hydrates is also strongly influenced by radiative processes, our modeled surface temperatures can be used to assess their thermal stability. For instance, comparing the dehydration timescales of hydrated materials such as epsomite, natron, and mirabilite \citep[see Figure~5 of][]{McCord2001} with the range of temperatures expected from our simulations shows that, despite exhibiting high temperatures, epsomite and natron could still be thermally stable at Europa's equator. Mirabilite, on the other hand, should not be thermally stable at this location given its lower dehydration timescale \citep[$10^3$~years vs. $>10^6$~years for epsomite and natron,][]{McCord2001}, but could persist at higher ($\geq 50$\textdegree) latitudes where maximum temperatures remain below 100~K (Figure~\ref{fig:TminTmax}b). Similarly, as for water ice, high daytime temperatures should promote the thermal regeneration of irradiation-amorphized sulfuric acid hydrates. Notably, \citet{Loeffler2012} show that for temperatures higher than 86~K, commonly reached across most of Europa's surface (Figure~\ref{fig:TminTmax}b), thermally driven regeneration of H$_2$SO$_4 \cdot $4H$_2$O and H$_2$SO$_4 \cdot $H$_2$O should be efficient over geological timescales. Future coupling between our numerical models and estimates of radiolytic destruction rates could better assess the hydration and crystalline state of Europa's surface materials.

\paragraph{Implications for future temperature measurements and the search for thermal anomalies by JUICE and Europa Clipper:}
The Europa Clipper and JUICE missions are expected to arrive in the Jovian system in 2030 and 2031, respectively, providing new measurements from E-THEMIS and SWI to advance our understanding of the thermodynamics of Europa icy surface and shallow subsurface. In particular, E-THEMIS will provide new, high-resolution measurements of Europa's surface temperatures. E-THEMIS will map more than 80\% of Europa's surface at multiple times of day at a resolution of 8~km per pixel \citep{Christensen2024}, offering an unprecedented dataset to complement this study and refine the determination of surface thermophysical properties. One of the primary science objectives of E-THEMIS is the detection of thermal anomalies indicative of recent or ongoing geologic activity, such as endogenic heat sources or active venting \citep{Christensen2024}. However, identifying such anomalies requires an accurate knowledge of the background surface temperature field driven by insolation and thermophysical properties. The thermal maps derived in this study therefore provide a critical baseline against which future E-THEMIS and SWI (Submillimetre Wave Instrument) observations can be compared.

To this end, we computed (Figure~\ref{fig:TminTmax_ETHEM}) the range of minimum and maximum surface temperatures expected over the nominal Europa Clipper mission duration \citep[through 2034,][]{Pappalardo2024}. Surface temperatures during the mission are expected to range between 67.6 and 141.2~K at the equator, with minimum nighttime temperatures between 67.6 and 93.1~K, and maximum daytime temperatures between 90.3 and 141.2~K. Notably, minimum temperatures are not expected to differ significantly from the annual minima derived above, whereas maximum temperatures are systematically lower. This is a consequence of Europa Clipper arriving near Europa during a period of minimum insolation, reducing peak daytime heating relative to the annual average. Deviations of E-THEMIS or SWI measurements from the temperature ranges predicted here could thus be used to identify regions of anomalous endogenic activity, constrain subsurface heat fluxes, or reveal localized variations in surface thermophysical properties not captured by current disk-integrated observations. Note that the detection of endogenic heat flux will be most tractable at high latitudes, where insolation-driven temperatures are lowest \citep{Christensen2024}.

\section{Conclusion \label{sec:Cl}}
In this study, we reanalyzed the Galileo PPR dataset to derive the surface properties of Europa, particularly albedo, thermal inertia, and microphysical properties at the spatial resolution of 0.5\textdegree (14~km at the equator) representing 22\% of the total surface. The main conclusions of our investigation are:

\begin{itemize}
\item The mean value of Europa's Bond albedo is 0.64~$\pm$~0.06 (1$\sigma$) (Figure~\ref{fig:albedo_timap}a), with significant longitudinal variations characterized by a darker trailing hemisphere and a brighter leading hemisphere.

\item The thermal inertia of Europa's surface (Figure~\ref{fig:albedo_timap}b) ranges from 18 to 150~\tiu, with a mean value of 56~$\pm$~17 (1$\sigma$). Significant spatial variations are observed, with a band of low thermal inertia in Europa's equatorial regions compared to the higher thermal inertia at mid-latitudes, as well as higher thermal inertia in the equatorial region of the trailing hemisphere compared to the leading hemisphere.

\item The thermal inertia of Europa's surface is significantly lower than that of solid, compacted water ice, consistent with a porous icy regolith. Assuming that the ice is purely crystalline water ice within the first centimeters of the subsurface, consistent with near-infrared spectroscopic studies \citep{Hansen2004,Ligier2016,Cartwright2025}, and applying the thermal conductivity models of \cite{Ferrari2016}, we conclude that the surface is composed of porous ice with a tight contact model between grains, with grain sizes ranging from a few $\mu$m to a few centimeters (Section~\ref{sec:resultsgrainsize}), and with a high average porosity of 0.61~$\pm$~0.1 (1$\sigma$) (Figure~\ref{fig:porositymap}).

\item With the exception of the Pwyll ejecta, which present lower albedo and higher thermal inertia, no significant correlation between thermal inertia and geological units as defined by \cite{Leonard2024} is identified on Europa (Figure~\ref{fig:geologicalunit}). This conclusion is consistent with previous studies by \cite{Rathbun2014}, \cite{Trumbo2017}, \cite{Thelen2024}, and \cite{Howes2025}.

\item Contrary to Saturn's icy moons Mimas, Tethys, and Dione, the equatorial regions of Europa do not exhibit a PacMan anomaly \citep[i.e., a 'lens-shaped' region of high thermal inertia, most likely created by electron-induced sintering;][]{Howett2011}. Energetic electrons are predicted to bombard the equatorial regions of both hemispheres, preferentially the trailing one \citep{Nordheim2018}. Yet the leading equator shows the lowest thermal inertia across the surface studied here (Figure~\ref{fig:albedo_timap}b), and the higher thermal inertia of the trailing equator is more likely due to compositional differences (Section~\ref{ssec:traillead}) than a signature of electron-driven sintering. Furthermore, the microphysical conditions required for electron-induced sintering are not satisfied (Section~\ref{ssec:pacman}).

\item The absence of large grain-size ice at Europa's equator may indicate that thermally driven sintering produces larger crystals at depth rather than at the surface. This hypothesis is supported by strong near-surface temperature gradients induced by the low thermal inertia (and hence low conductivity) of the surface, potentially triggering temperature-gradient metamorphism, which produces depth hoar on Earth \citep{Colbeck1983} and has been proposed as a significant sintering agent on icy moons by \cite{Mergny2026}. Further modeling and laboratory experiments are required to investigate this mechanism under extraterrestrial conditions.

\item The correlation between the higher thermal inertia observed at high latitudes (Figures~\ref{fig:albedo_timap}b and \ref{fig:thermalinertiavslatitude}) and the spatial distribution of magnetospheric ion fluxes from \cite{Nordheim2022} suggests that ion sputtering-driven sintering may be a major agent controlling sintering on Europa, as proposed by \cite{Clark1983}. However, the detailed physical mechanisms involved remain poorly understood. Additional modeling and laboratory experiments are needed to investigate this process under Europa-like conditions.

\item Differences in thermal inertia at the equator between the leading and trailing hemispheres likely reflect compositional differences between the two hemispheres. The trailing equatorial region is mostly composed of sulfuric acid hydrates \citep{Ligier2016,King2022}, whereas the leading hemisphere is dominated by water ice \citep{Ligier2016} (Section~\ref{ssec:traillead}). Sulfuric acid hydrates may act as a cementing agent, increasing the effective conductivity of the porous ice compared to clean porous ice. They are also likely responsible for the lower albedo of the trailing side.

\item No significant vertical layering is detected in the shallow subsurface structure (Figure~\ref{fig:layering}), as diurnal PPR measurements are best reproduced by a uniform thermal inertia structure within the diurnal skin depth (first centimeter). However, comparison between the porosity derived in this study and values inferred from eclipse observations in the first millimeters of the surface \citep{Hansen1973}, as well as ALMA observations probing depths of 10–20~cm \citep{Thelen2024} and models \citep{Mergny2024compaction}, suggests gradual compaction with depth over the upper 10–20~cm, most likely resulting from sintering processes (thermally-driven or sputtering-driven) (Section~\ref{ssec:layering}).

\item Using the thermophysical properties derived in this study, we estimate that surface temperatures at mid to low latitudes on Europa range between 67.5~K and 148~K (Figure~\ref{fig:TminTmax}). While the highest temperatures at the equator do not favor the stability of volatiles at the surface, peak daytime temperatures promote the thermal (re)crystallization of water ice and acid hydrates, but are not high enough to efficiently dehydrate hydrated minerals such as epsomite and natron. However, these thermally driven processes likely compete with radiative processes, highlighting the need for a better understanding of their relative roles.

\item Surface temperatures at mid to low latitudes are expected to range between 67.6 and 141.2~K during the JUICE and Europa Clipper flybys between 2031 and 2034 (Figure~\ref{fig:TminTmax_ETHEM}). Deviations from these predicted temperature ranges in future E-THEMIS and SWI observations could indicate localized endogenic activity, constrain subsurface heat fluxes, and/or reveal small-scale variations in surface thermophysical properties not captured by current observations.
\end{itemize}

The Europa Clipper and JUICE missions are expected to arrive in the Jovian system in 2030 and 2031, respectively, and will provide unique, high-resolution, and comprehensive datasets to improve our understanding of this system. In the meantime, our study highlights the need for an improved understanding of the physics of ice and of the competition between thermally driven and radioactively driven processes on Europa, through both modeling and laboratory experiments.

\begin{acknowledgements}
    The authors thank Erin J. Leonard for providing Europa geological map. 
      LL's research was supported by an appointment to the NASA Postdoctoral Program  administered by Oak Ridge Associated Universities  at the Jet Propulsion Laboratory,  California Institute of Technology, under a contract with the National Aeronautics and Space Administration (80NM0018D0004). Part of this work was performed at at the Jet Propulsion Laboratory, California Institute of Technology, under a contract with the National Aeronautics and Space Administration (80NM0018D0004).  Some of the computational analyses were run on Northern Arizona University's Monsoon computing cluster, funded by Arizona's Technology and Research Initiative Fund. © 2026. All rights reserved.

      PPR data are available on the Planetary Data System \cite{GalileoPDS}. The thermal model, KRC, used in this study can be retrieved at: \url{https://krc.mars.asu.edu/}. Data files used to generate the figures in this analysis are available in a public repository, see \cite{Lange2026europadata}. 
\end{acknowledgements}

\bibliographystyle{aa} 
\bibliography{europa} 
\onecolumn

\begin{appendix}
\section{Thermal inertia of amorphous and crystalline ice}

    \begin{figure*}[h!]
    \sidecaption
 \includegraphics[width=0.7\textwidth]{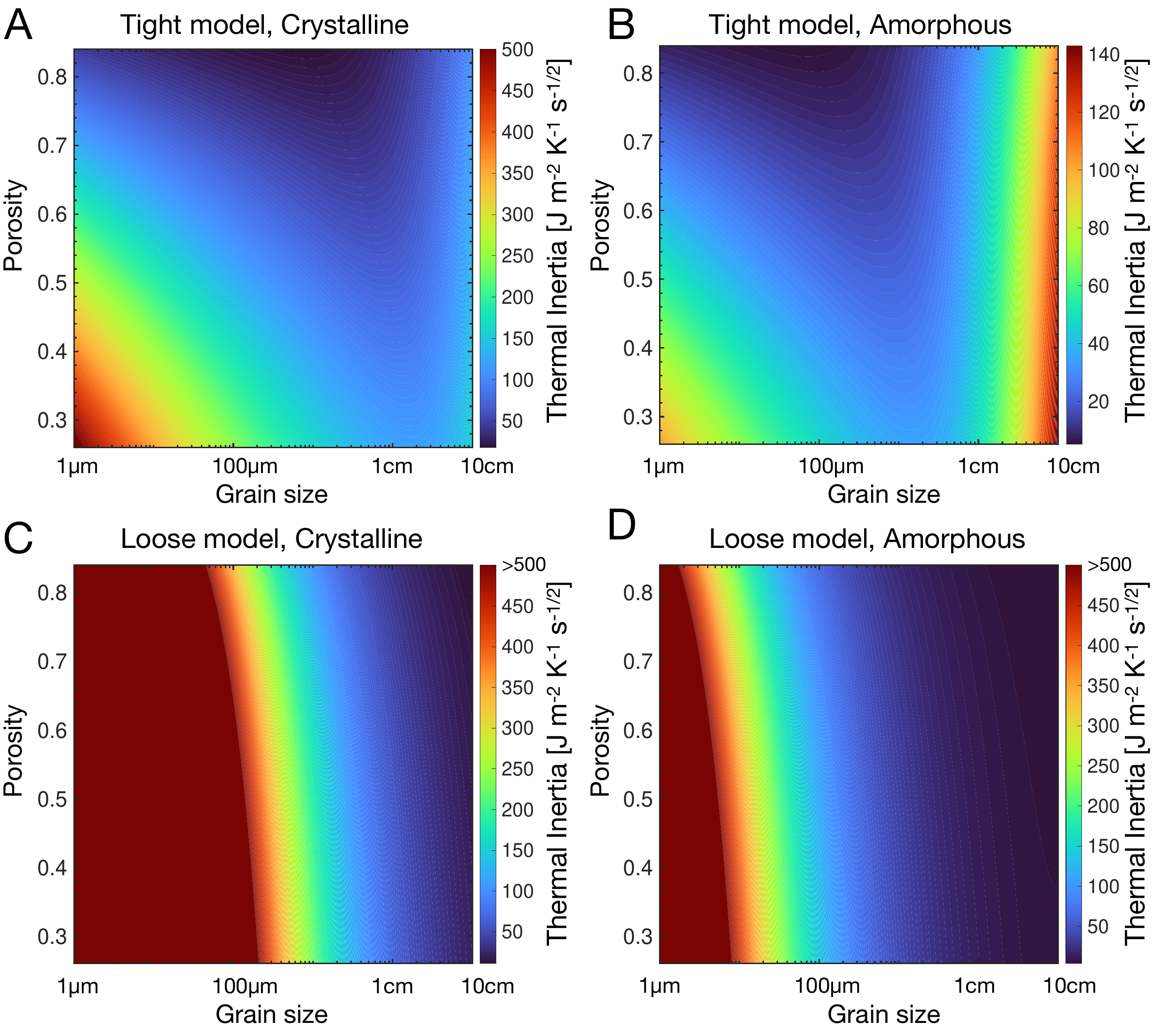} \caption{Thermal inertia as a function of the ice grain size and porosity, for crystalline  (left) and amorphous (right) ice for the tigh (up) and loose (down) model.} 
    \label{fig:thermalinertia_theo_allice}
    \end{figure*}

\section{Retrieved ice micro-physical properties assuming amorphous ice}

    \begin{figure*}[h!] 
        \sidecaption
 \includegraphics[width=0.7\textwidth]
    {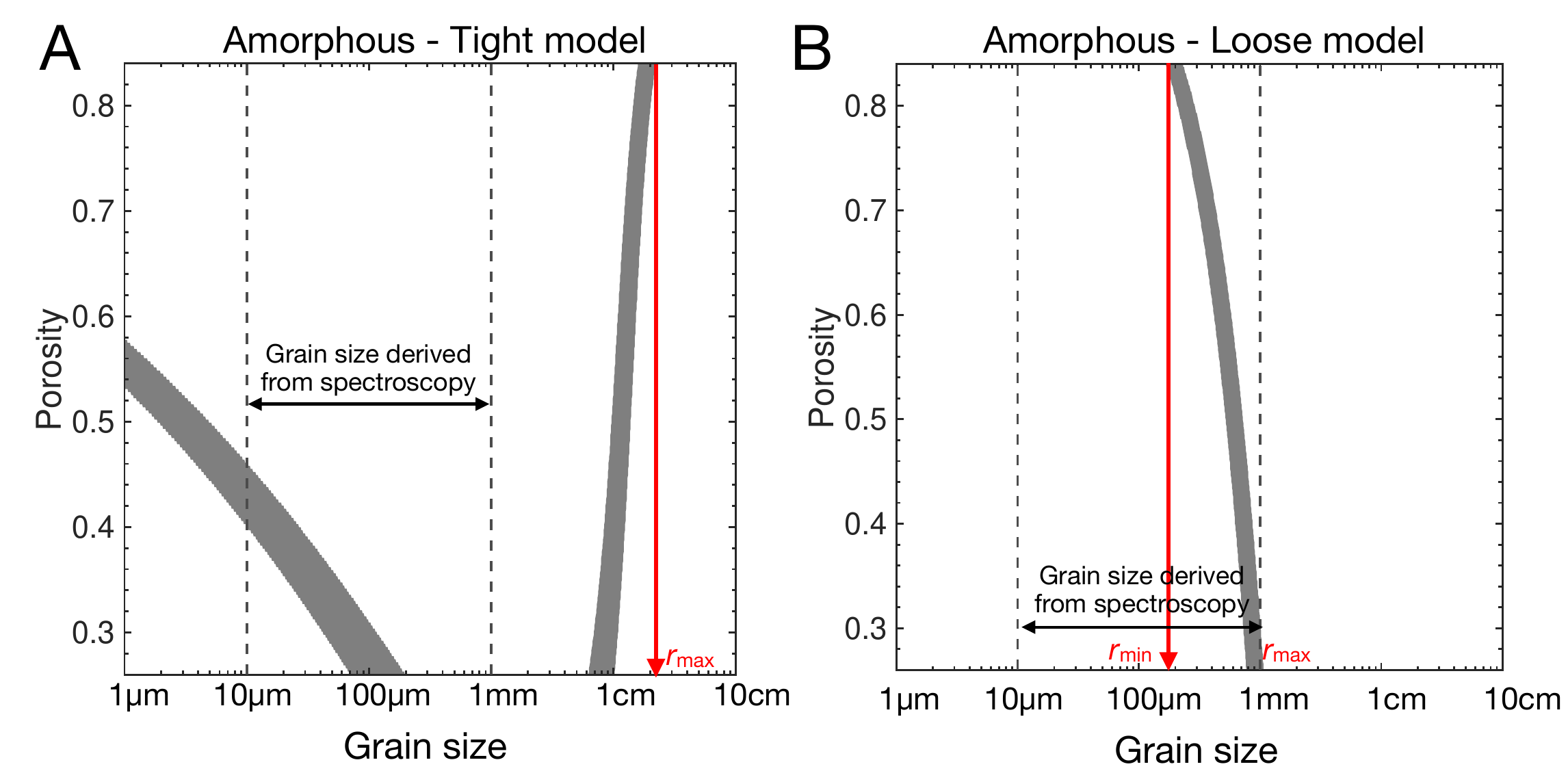} \caption{Same as Figure~\ref{fig:illustrationmethod} (panels c and d), but assuming amorphous ice instead  of crystalline ice.  Grain size and porosity combinations consistent with the derived  thermal inertia (and its associated uncertainty) are shown in grey for the tight (a) and loose (b)  models. The ranges of admissible  grain sizes ($r_{\rm min}$ and $r_{\rm max}$) and the minimum porosity  ($p_{\rm min}$) are derived following the same procedure described in  Section~\ref{ssec:derivingrandp}.} 
    \label{fig:illustrationmethod_amorphous}
    \end{figure*}

\newpage
\section{Effect of solar radiation absorption within the ice on surface temperature}

\begin{figure*}[h!]
    \sidecaption
\includegraphics[width=0.5\textwidth]{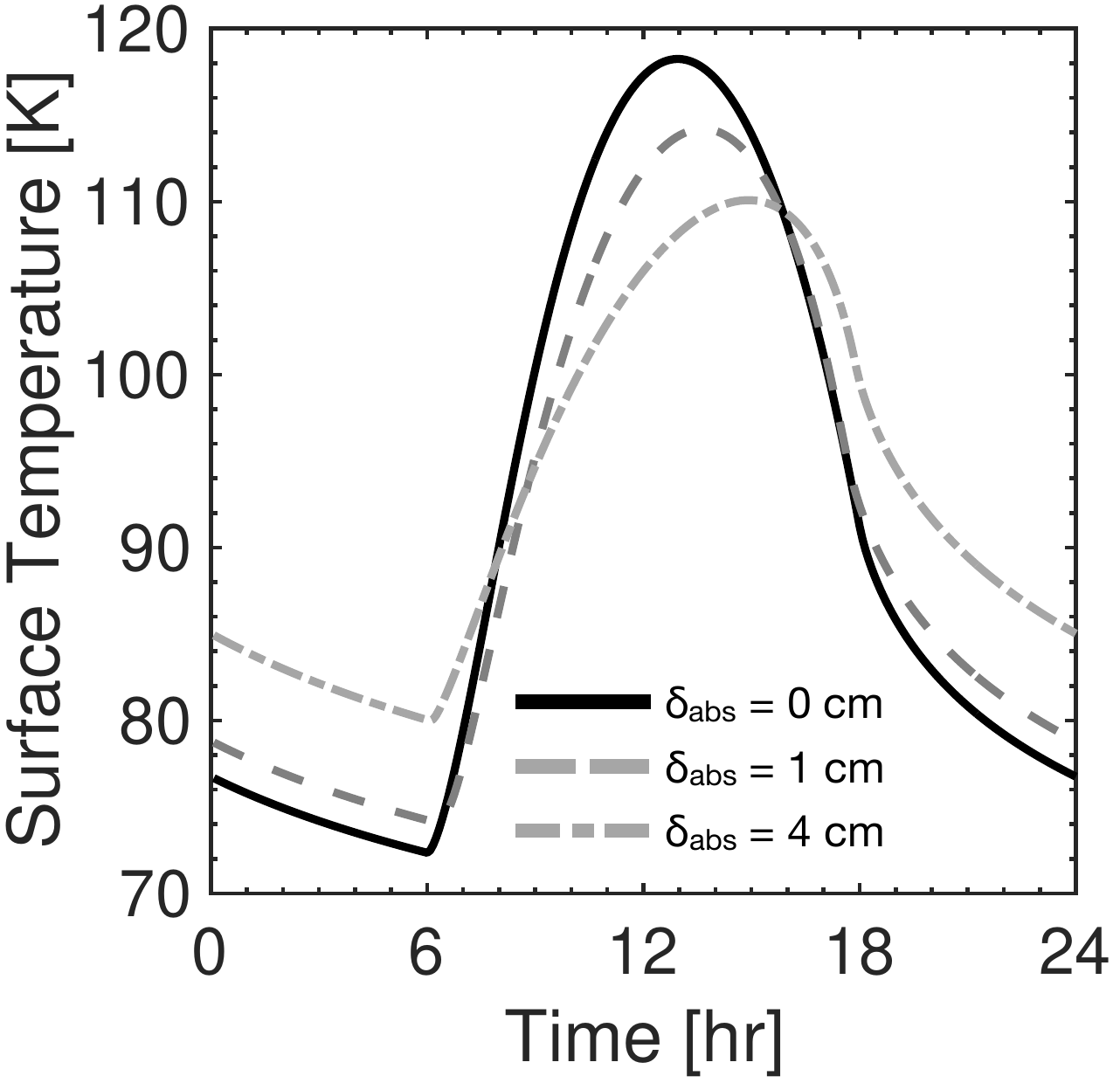}
\caption{
Surface temperatures modeled with KRC assuming volumetric absorption of solar radiation within the subsurface, parameterized by an  e-folding absorption depth $\delta_{\rm{abs}}$. The black curve corresponds to complete absorption at the surface  ($\delta_{\rm{abs}}$~=~0~cm), the gray dashed curve to  $\delta_{\rm{abs}}$~=~1~cm, and the gray dash–dotted curve to $\delta_{\rm{abs}}$~=~4~cm. Simulations were performed at the equator, assuming a thermal inertia of 33~\tiu\ and an albedo of 0.72.}
\label{fig:illustration_solarabsinfround}
\end{figure*}

\section{Impact of sub-pixel mixing on retrieved thermophysical properties \label{app:mixing}}
To assess the effect of sub-pixel mixing arising from albedo and thermal inertia heterogeneities, we follow the approach of \cite{PUTZIG2007a,PUTZIG2007b}. We consider two terrains, $S_1$ and $S_2$, with respective bond albedos $A_1$ and $A_2$ and thermal inertias $\mathrm{TI}_1$ and $\mathrm{TI}_2$. The areal fraction occupied by $S_1$ is $f_1$, and that of $S_2$ is $(1-f_1)$.

\begin{figure}
\centering
\includegraphics[width=\linewidth]{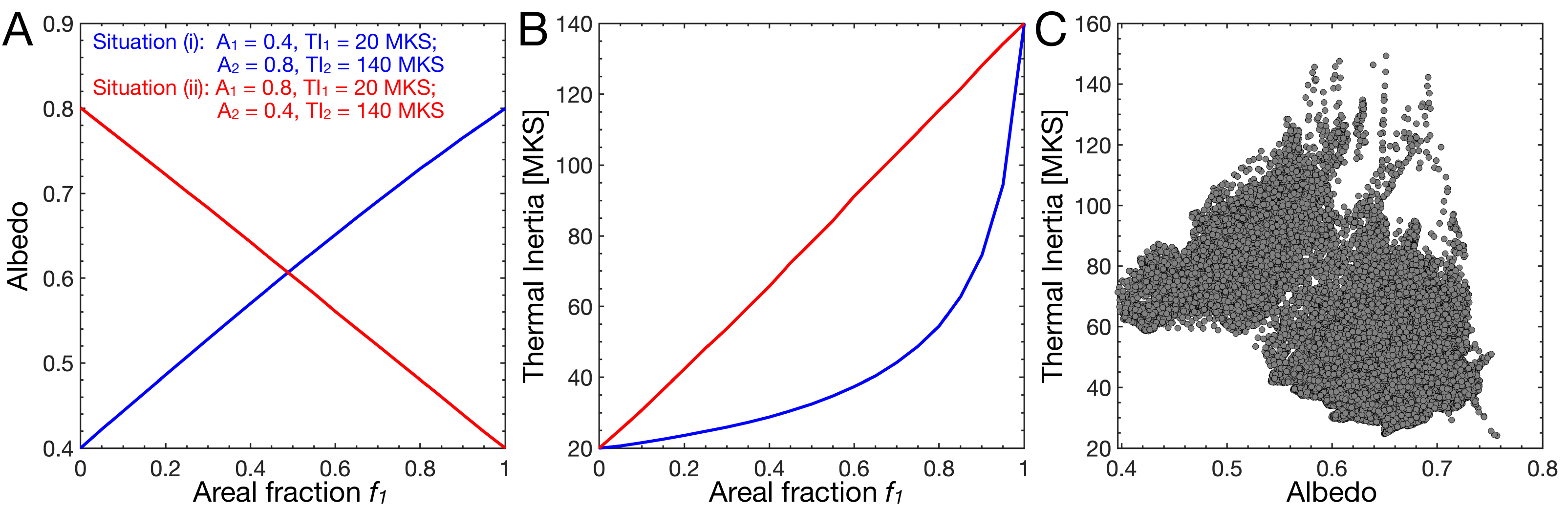}

\caption{Effect of sub-pixel mixing on the retrieved thermophysical properties. (a) Retrieved Bond albedo and (b) retrieved thermal inertia as a function of the fractional coverage of terrain $S_1$. The blue curve corresponds to case (i), while the red curve corresponds to case (ii). (c) Retrieved thermal inertia values from Section~\ref{sec:Results} plotted against the retrieved Bond albedo.}
\label{fig:subpixelmixing}
\end{figure}

For each terrain we compute the surface temperatures $T_{S1}$ and $T_{S2}$ at 13:00 and 20:00 local solar time, representative of the retrievals used in this study. The equivalent pixel
surface temperature $T_\mathrm{surf,eq}$ is obtained by equating radiances, assuming each terrain radiates as a blackbody:
\begin{equation}
    T_\mathrm{surf,eq}^{4} = f_1~T_{S1}^{4} + (1-f_1)~T_{S2}^{4}
    \label{eq:mixing}
\end{equation}
From these temperatures, we retrieve the equivalent pixel albedo and thermal inertia following the method described in Section~\ref{ssec:constrainAlbTI}. We performed two test cases: (i)~$S_1$ is a low-albedo ($A_1~=~0.4$), low-thermal-inertia ($\mathrm{TI}_1~=~20$~\tiu) terrain, and $S_2$ a high-albedo ($A_2~=~0.8$), high-thermal-inertia ($\mathrm{TI}_2~=~140$~\tiu) terrain (these values bracketing
the measured albedo and thermal-inertia ranges, see Section~\ref{sec:Results}); (ii)~$S_1$ is a high-albedo ($A_1~=~0.8$), low-thermal-inertia ($\mathrm{TI}_1~=~20$~\tiu) terrain, and $S_2$ a low-albedo ($A_2~=~0.4$), high-thermal-inertia ($\mathrm{TI}_2~=~140$~\tiu) terrain. The results are presented in Figure~\ref{fig:subpixelmixing}a,b.

For the albedo (Figure~\ref{fig:subpixelmixing}a), both cases show that it is difficult to isolate the effect of sub-pixel mixing on the retrieved effective albedo, the latter varying linearly with $f_1$. This behaviour is expected: neglecting conduction, the radiative cooling of the surface scales
as (using the notation of Eq.~\ref{eq:GeneralTsurf}):
\begin{equation}
    T_\mathrm{surf,eq}^{4} \propto \left[ f_1~(1-A_1) + (1-f_1)~(1-A_2) \right] F_\mathrm{solar}
\end{equation}
so that the retrieved equivalent albedo satisfies:
\begin{equation}
1-A_\mathrm{eff} = f_1~(1-A_1) + (1-f_1)~(1-A_2)
\end{equation}
i.e.,
\begin{equation}
    A_\mathrm{eff} = f_1~A_1 + (1-f_1)~A_2
\end{equation}

For the thermal inertia (Figure~\ref{fig:subpixelmixing}b), the outcome depends on the case considered. In case~(i), the combination of low albedo and low thermal inertia in $S_1$ produces a strong temperature contrast with $S_2$ (high albedo, high thermal inertia), particularly during the daytime. Because the pixel-integrated radiance scales as $T^{4}$, the hotter terrain ($S_1$) dominates the daytime signal and, consequently, the retrieval. As $S_1$ becomes the minority terrain, the measured signal is progressively dominated by $S_2$. In this configuration, the retrieved thermal inertia varies nonlinearly (with a markedly convex shape) as a function of $f_1$, allowing sub-pixel mixing to be constrained.

In case~(ii), a compensation effect arises: $S_1$ has a high albedo (and is therefore cold during the day) and a low thermal inertia (and thus cools rapidly at night), whereas $S_2$ has a low albedo (warm during the day) and a high thermal inertia (slow cooling at night). The terrain with the largest diurnal temperature amplitude (here $S_2$) is also the warmest on average. As $f_1$ increases (i.e., as the fraction of the colder, lower-contrast terrain $S_1$ increases), the contribution of the warmer terrain $S_2$ decreases steadily, because both the mean temperature and the diurnal amplitude vary in the same direction. This results in an almost linear variation of the retrieved thermal inertia with $f_1$. In this case, the physical interpretation is degenerate: the retrieved thermal inertia can be reproduced by an infinite family of sub-pixel configurations.

The distribution of retrieved albedo and thermal inertia from PPR data (Figure~\ref{fig:subpixelmixing}c) is highly dispersed. . Excluding a few outliers (terrains with an albedo lower than 0.45 and terrains with a thermal inertia higher than 120~\tiu, both observed on the trailing hemisphere), a slight anti-correlation between albedo and thermal inertia is apparent in the bulk of the distribution (Pearson coefficient $r~=~-0.46$). For instance, terrains with thermal inertia in the range 80~-~100~\tiu~exhibit a median albedo of 0.54~$\pm$0.06 (1$\sigma$), whereas those in the range 20~-~40~\tiu~have a median albedo of 0.66~$\pm$0.04 (1$\sigma$). This configuration of higher thermal inertia co-occurring with lower albedo corresponds to the configuration described in case~(ii). Consequently, the physical interpretation of our retrieval should be degenerate: the retrieved albedo and thermal inertia values are compatible with an infinite family of sub-pixel configurations.

\section{Correlation between thermal inertia and modeled magnetospheric ion fluxes}

\begin{figure}[h!]
\centering
\includegraphics[width=0.9\linewidth]{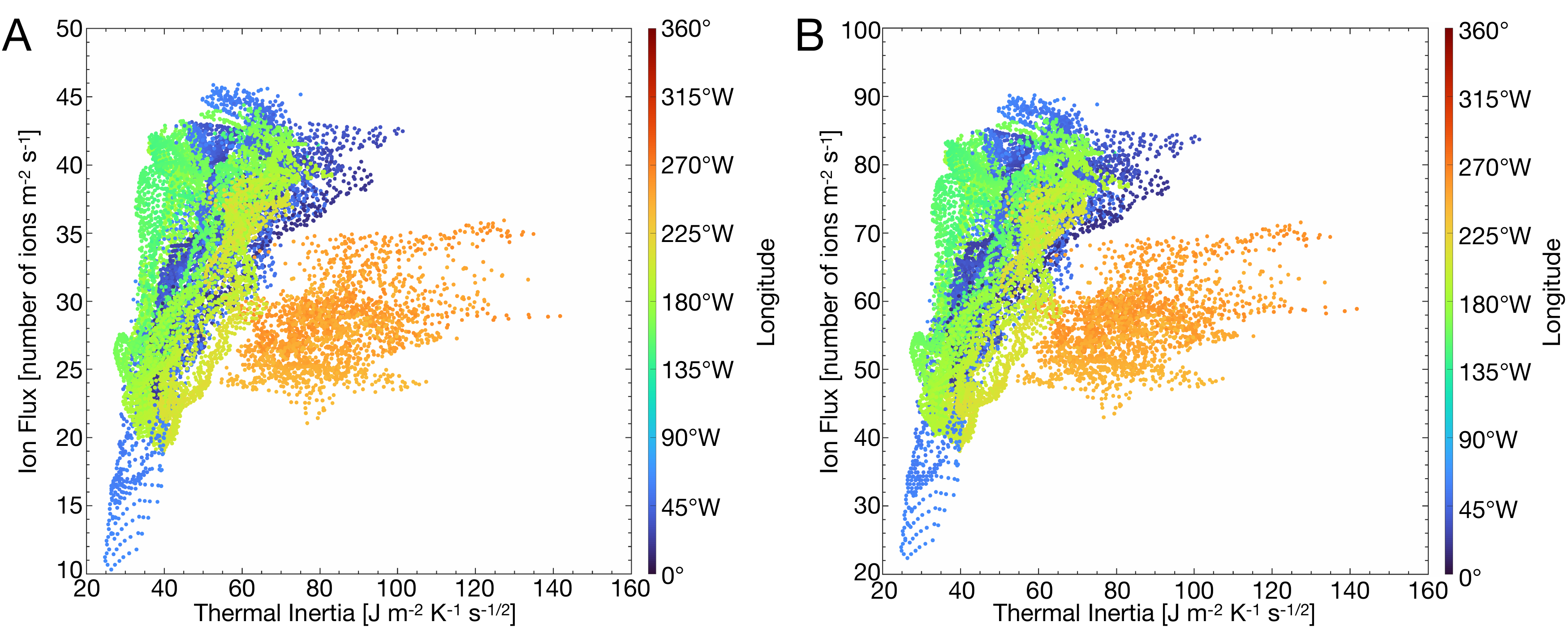}

\caption{Modeled (a) oxygen and (b) sulfur ion fluxes of \cite{Nordheim2022} as a function of the thermal inertia retrieved in this study.  Points are colored by west longitude.}
\label{fig:appendix:magnetosphericflux}

\end{figure}

\newpage
\section{Range of surface temperatures expected during the Europa Clipper mission}

\begin{figure}[h!]
\centering
\includegraphics[width=\linewidth]{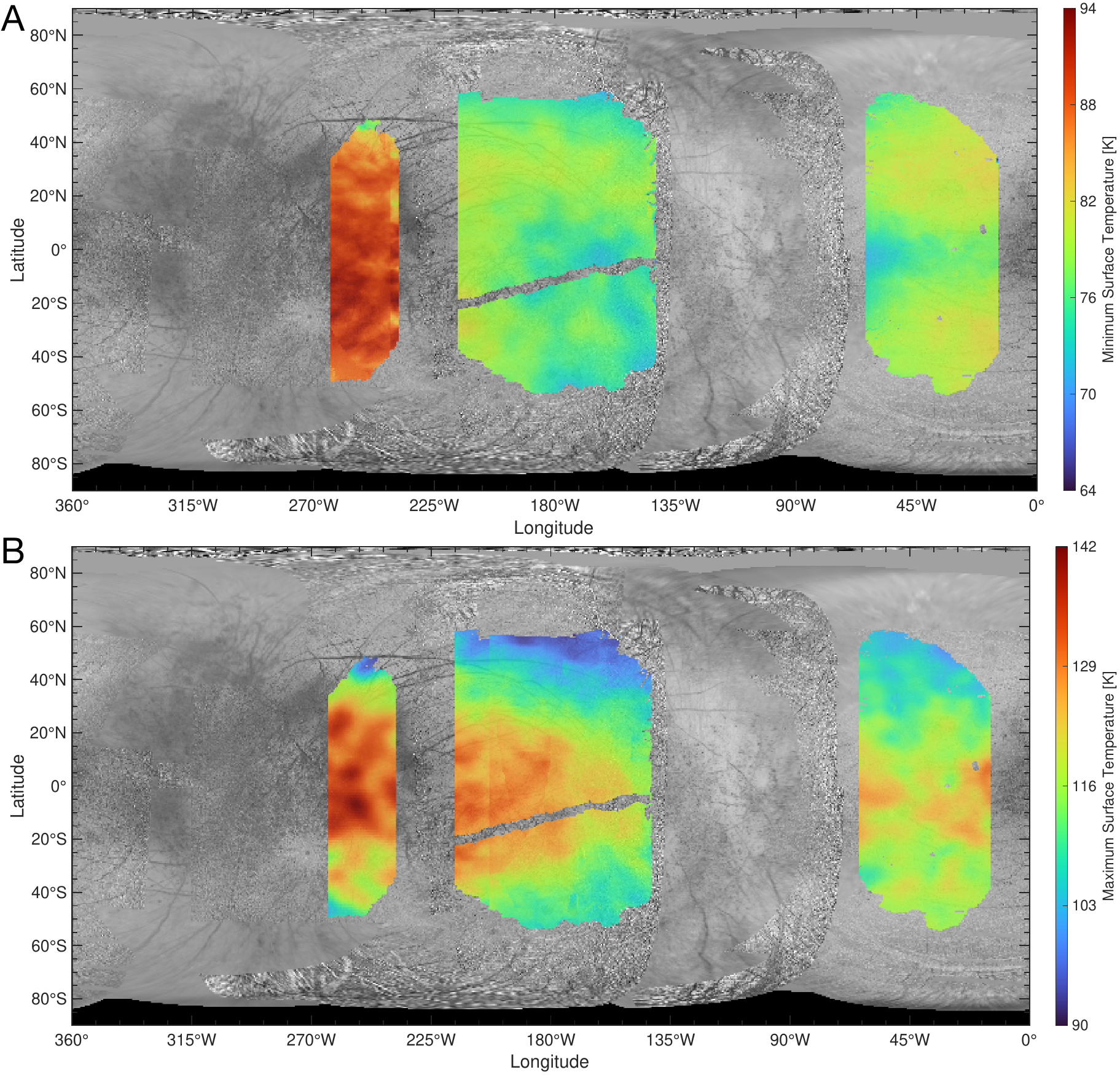}

\caption{Minimum (a) and maximum (b) surface temperatures expected between 2030 and 2034, during the Europa Clipper mission \citep{Pappalardo2024}. Temperatures are simulated with KRC using the albedo and thermal inertia maps generated in Figure~\ref{fig:albedo_timap}. An emissivity of 0.9 is assumed.}
\label{fig:TminTmax_ETHEM}
\end{figure}
\end{appendix}

\end{document}